\documentclass[11pt]{article}

\usepackage[margin=1in]{geometry}
\usepackage[T1]{fontenc}
\usepackage[utf8]{inputenc}
\usepackage{lmodern}
\usepackage{amsmath,amssymb}
\usepackage{graphicx}
\usepackage{booktabs}
\usepackage{tabularx}
\usepackage[ruled,vlined]{algorithm2e}
\usepackage{xcolor}
\usepackage{hyperref}
\usepackage{amsthm}

\newtheorem{example}{Example}

\newcommand{\benchgen}{\textsc{BenchGen}}

\begin{document}

\title{Multi-Language Benchmark Generation via L-Systems}

\author{
Vinicius Francisco da Silva \\
UFMG, Brazil \\
\texttt{silva.vinicius@dcc.ufmg.br}
\and
Heitor Leite \\
UFMG, Brazil \\
\texttt{heitor.leite@dcc.ufmg.br}
\and
Fernando Magno Quint\~ao Pereira \\
UFMG, Brazil \\
\texttt{fernando@dcc.ufmg.br}
}

\date{}

\maketitle

\begin{abstract}
L-systems are a mathematical formalism proposed by biologist Aristid Lindenmayer with the aim of simulating organic structures such as trees, snowflakes, flowers, and other branching phenomena. They are implemented as a formal language that defines how patterns can be iteratively rewritten. This paper describes how such a formalism can be used to create artificial programs written in programming languages such as C, C++, Julia and Go. These programs, being large and complex, can be used to test the performance of compilers, operating systems, and computer architectures. This paper demonstrates the usefulness of these benchmarks through multiple case studies. These case studies include a comparison between \texttt{clang} and \texttt{gcc}; a comparison between C, C++, Julia and Go; a study of the historical evolution of \texttt{gcc} in terms of code quality; a look into the effects of profile guided optimizations in \texttt{gcc}; an analysis of the asymptotic behavior of the different phases of \texttt{clang}'s compilation pipeline; and a comparison between the many data structures available in the Gnome Library (\textsc{GLib}).
These case studies demonstrate the benefits of the L-System approach to create benchmarks, when compared with fuzzers such as \textsc{CSmith}, which were designed to uncover bugs in compilers, rather than evaluating their performance.
\end{abstract}

\noindent\textbf{Keywords:}
L-System, Benchmark, Synthesis

\maketitle

\section{Introduction}
\label{sec_intro}

Language processing systems, such as compilers, interpreters and static analyzers are complex tools whose validation depends on programs written in the target language. 
However, the number of available benchmarks for any given compiler is typically limited~\cite{Wang18}. 
To address this limitation, several tools have been developed to automatically generate test programs~\cite{Yang11}. 
This process, known as \emph{fuzzing}~\cite{Manes21}, is widely used to uncover bugs such as crashes and memory leaks. 
In the compiler domain, fuzzers such as Csmith~\cite{Yang11} and YARPGen~\cite{Livinskii20} generate random C programs to stress-test compilers and support static analysis. 
More recently, fuzzers like Fuzz4All~\cite{Xia24} have leveraged Large Language Models (LLMs) to generate test programs not only for compilers but also for constraint solvers, interpreters, and other software systems with accessible APIs.

Despite the abundance of program generators~\cite{Yang11}, existing tools exhibit important limitations. 
A key issue is the lack of control over the size of the generated programs. 
For example, Csmith, the most widely used C compiler fuzzer, does not allow users to tune the output size. 
Instead, it produces programs whose sizes follow a normal distribution. 
When compiled with \texttt{clang} v9.0.1 at \texttt{-O0}, these programs contain on average 20,190 LLVM instructions, with a standard deviation of 3,650 and a median of 19,161 instructions~\cite{Faustino21}. 
Other fuzzers, such as YARPGen~\cite{Livinskii20}, LDRGen~\cite{Barany17}, and Orange3~\cite{Kitaura18}, show similar behavior. 
As a consequence, these tools are ill-suited for performance evaluation of compiler components such as parsing, semantic analysis, and code generation.

\paragraph{Programs via L-Systems. }
To overcome this limitation, this paper proposes a methodology for stress-testing compilers with a focus on \emph{performance} rather than solely on \emph{correctness}. 
Our approach enables the controlled generation of programs whose size can be precisely tuned by the user. 
This makes it possible to construct synthetic code of virtually arbitrary size, constrained only by practical resources such as generation time and storage space.

The central insight is that programs often exhibit recursive, self-similar structures, as discussed in Section~\ref{sec_obf}. 
For instance, the branches of a control construct (e.g., \texttt{if-then-else}) are themselves programs that may recursively contain further control constructs. 
To exploit this property, we introduce a generation technique based on \emph{L-systems}~\cite{Lindenmayer68}. 
Originally developed by Aristid Lindenmayer to model the growth of biological organisms, L-systems provide a grammar-based framework that we extend to code generation. 
This paper shows how families of self-similar programs can be encoded as L-grammars. 
Programs are then generated by iterative rewriting, where each resulting string encodes the blueprint of a program organized around a central data structure such as an array or a list. 
As discussed in Section~\ref{sec_sol}, this technique supports multiple programming languages, yields complex control-flow graphs, avoids undefined behavior, and enables the manipulation of different data structures.

\paragraph{Summary of Contributions. }
To demonstrate the benefits of representing program growth with L-systems, we introduce \benchgen, a multi-language benchmark generator. 
Section~\ref{sec_eval} evaluates \benchgen{} through seven case studies, including: a comparison between \texttt{gcc} and \texttt{clang}; a comparison of Go, Julia, C, and C++; an asymptotic analysis of different components of \texttt{clang}; a longitudinal analysis of the evolution of \texttt{gcc}; an evaluation of profile-guided optimizations in \texttt{clang}; a comparison of data structures from \textsc{GLib}; and a comparison between programs generated by \benchgen{} and by \textsc{Csmith}. 
\benchgen{} is publicly available under the GPL 3.0 license, and has already been used to generate benchmarks for several languages beyond those discussed in this paper, including \textsc{Rust} and \textsc{Vale}.

\section{L-Systems}
\label{sec_obf}

An L-system (or {\it Lindenmayer system}) is a formal model based on rewriting rules, originally devised to describe the growth patterns of plants and other fractal-like structures. It comprises an alphabet of symbols, a set of production rules that define how symbols are transformed, and an initial \textit{string} (the axiom) that serves as the starting point. At each iteration, the rules are recursively applied to the current string, producing increasingly complex sequences. Example~\ref{ex_sierpinski} illustrates how this formalism operates.

\begin{figure}[t]
\centering
\includegraphics[width=0.7\columnwidth]{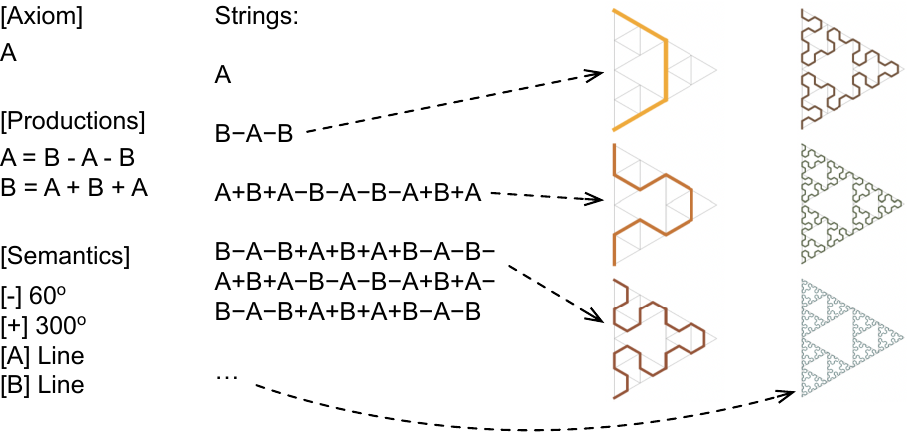}
\caption{L-system describing the Sierpinski Triangle.}
\label{fig_lSystem}
\end{figure}

\begin{example}
\label{ex_sierpinski}
Figure~\ref{fig_lSystem} presents an example of an L-system. The rules used in this system generate geometric patterns through string rewriting. Starting from the axiom $A$, the productions specify how symbols evolve at each step: $A \rightarrow B - A - B$ and $B \rightarrow A + B + A$. Here, the symbols $-$ and $+$ represent rotations of 60 and 300 degrees, respectively. Repeated application of these rules generates sequences that, when interpreted graphically, produce intricate fractal curves, such as the well-known Sierpinski Triangle.
\end{example}

\subsection{Programs as Self-Similar Structures}
\label{sub_autosimil}

L-systems exhibit a property known as {\it self-similarity}, meaning that structures contain smaller copies of themselves across different scales.
In the context of L-systems, this feature arises naturally from the recursive application of rewriting rules, producing patterns that preserve the same shape at progressively finer levels of detail. This behavior is evident in fractals like the Sierpinski Triangle seen in Example~\ref{ex_sierpinski}, where each component is a scaled-down replica of the whole.
Such hierarchical repetition is key to modeling phenomena like plant growth, coastlines, and tree branching.

Self-similarity also emerges in computer programs, which often embody recursive and hierarchical structures. Many programs are composed of smaller functions that may invoke themselves or be embedded within one another, as in \texttt{if-then-else} blocks or loops.
Modularity and code reuse further reinforce this pattern: generic routines can be instantiated repeatedly across different abstraction levels, as Example~\ref{ex_autoSimilarity} explains.

\begin{figure}[t]
\centering
\includegraphics[width=0.7\columnwidth]{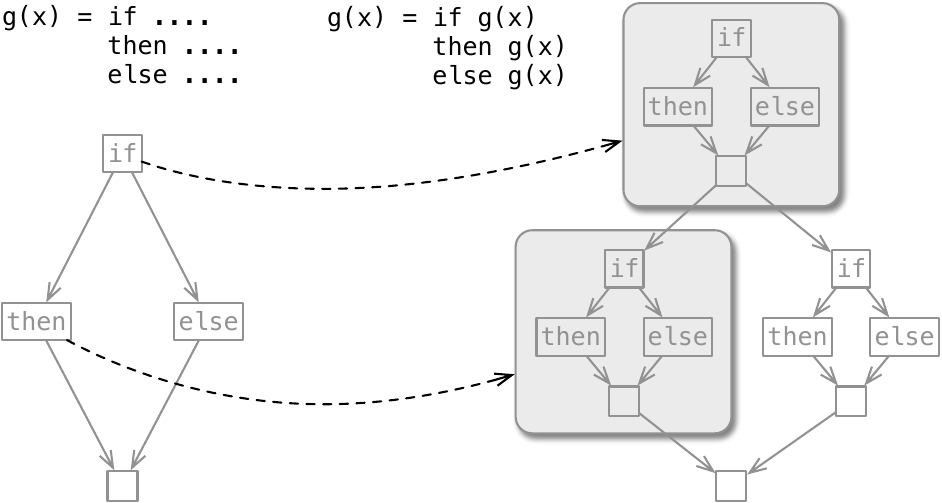}
\caption{The self-similar nature of computer code.}
\label{fig_autoSimilarity}
\end{figure}

\begin{example}
\label{ex_autoSimilarity}
Figure~\ref{fig_autoSimilarity} illustrates the concept of self-similarity in code using a nested \texttt{if-then-else} block. Initially, a function \texttt{g(x)} contains a single conditional. However, it can be recursively expanded to \texttt{g(x) = if g(x) then g(x) else g(x)}, forming a self-referential structure. Such recursive definitions naturally lead to self-similarity and are common in syntactical constructs that encode control-flow in programs.
\end{example}

\begin{figure}[t]
\centering
\includegraphics[width=1\columnwidth]{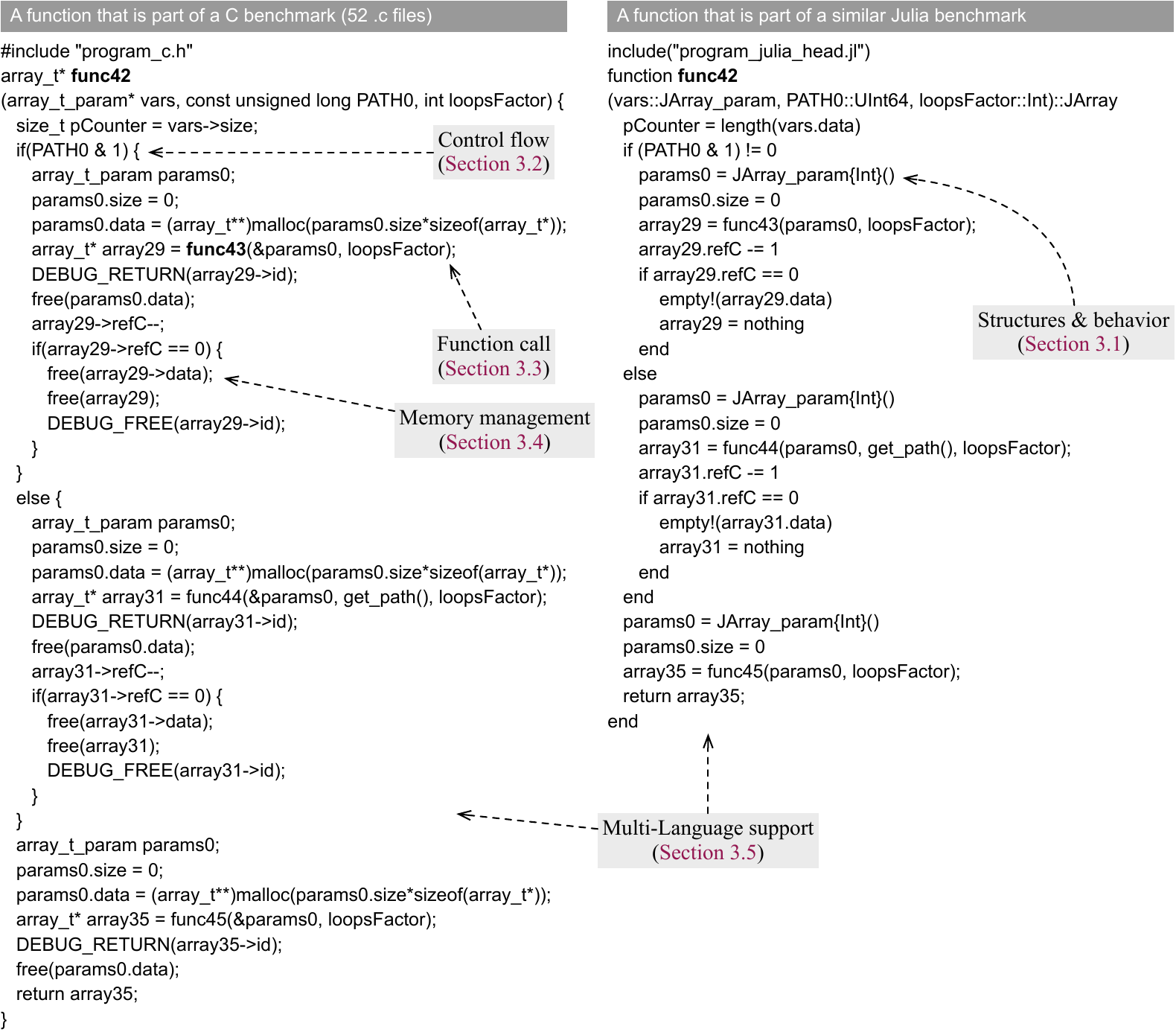}
\caption{Example of code that \benchgen{} produces for C and Julia that manipulate arrays.}
\label{fig_exampleC_Julia}
\end{figure}

\noindent
\textbf{Summary of Ideas}
This paper leverages the principle of self-similarity to generate C programs that are both well-defined and arbitrarily complex. The generation model is based on an L-grammar, akin to the one illustrated in Example~\ref{ex_sierpinski}, but instead of producing geometric patterns, it synthesizes C code constructions with executable semantics.
Figure~\ref{fig_exampleC_Julia} shows an example of two versions of a program that \benchgen{} produces for C and Julia.
This file is part of a 52-file benchmark that we use in Section~\ref{sub_multilang_comparison} to compare the performance of different language processing systems.

\section{Code Generation via L-Systems}
\label{sec_sol}

\begin{figure}[t]
\centering
\includegraphics[width=1\columnwidth]{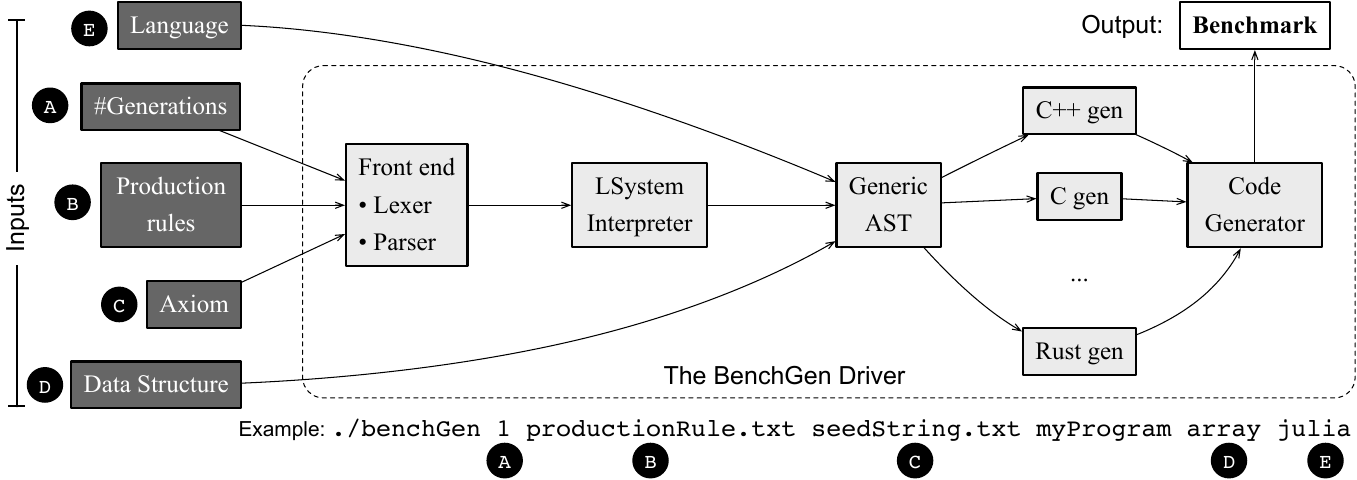}
\caption{\benchgen{} is a tool that generates programs in different programming languages. It receives as input the description of an L-System, which consists of a set of production rules, plus a starting symbol (the axiom). It interprets these rules, producing strings (the L-Strings), which guide the construction of genertic ASTs. These trees can be converted to programs in different programming languages using different containers. A code generator then converts the language-specific tree to a program.}
\label{fig_overview}
\end{figure}

The tool \benchgen{}, developed in this work, generates programs that manipulate data structures, according to the schema seen in Figure~\ref{fig_overview}. Section~\ref{sub_buildingBlocks} introduces the core building blocks used in program construction, while Section~\ref{sub_controlFlow} describes semantics of these building blocks.

\subsection{Syntactical Building Blocks}
\label{sub_buildingBlocks}

The L-systems described in this paper are built from two families of constructs:

\begin{itemize}
\item \textbf{Structure:} elements that define the control flow of a program, including \texttt{IF}, \texttt{LOOP}, and \texttt{CALL}.
\item \textbf{Behavior:} operations that specify how data is manipulated, including \texttt{new}, \texttt{insert}, \texttt{remove}, and \texttt{contains}.
\end{itemize}

\subsubsection{Structure Blocks}
\label{sss_estrutura}

The control flow in programs generated by \benchgen{} arises from combining four types of code blocks, as specified by the grammar below:

\[
\begin{array}{rcll}
b &::=& \text{IF}(b_{\text{cond}}, b_{\text{then}}) & ;; \mathit{if\_then}\\
  &\mid& \text{IF}(b_{\text{cond}}, b_{\text{then}}, b_{\text{else}}) & ;; \mathit{if\_then\_else} \\
  &\mid& \text{LOOP}(b_{\text{cond}}, b_{\text{body}}) & ;; \mathit{while} \\
  &\mid& \text{CALL}(b) & ;; \mathit{function\_call}
\end{array}
\]

Each of these constructs corresponds to a familiar programming construct: conditional branches (\texttt{if-then}, \texttt{if-then-else}), loops (\texttt{while}), and function calls.
Example~\ref{ex_structure} provides an illustration.

\begin{figure}[t]
\centering
\includegraphics[width=0.7\columnwidth]{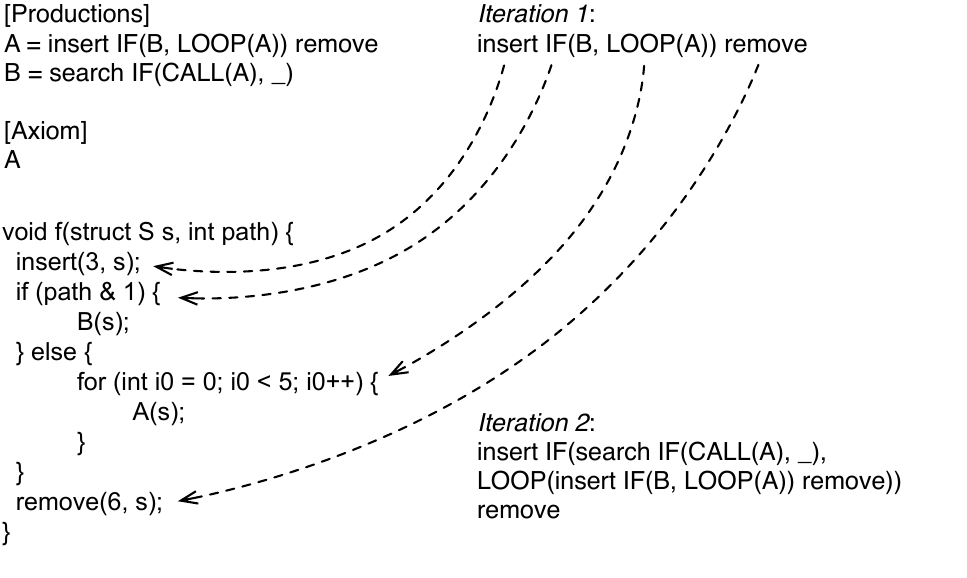}
\caption{Example of an L-grammar used to define programs.
The figure shows a simplified version of a C program.
Figure~\ref{fig_exampleC_Julia} provides a more concrete view.}
\label{fig_exBenchGrammar}
\end{figure}

\begin{example}
\label{ex_structure}
Figure~\ref{fig_exBenchGrammar} shows a grammar designed to synthesize programs. The right-hand side illustrates two derivation steps from this L-grammar, along with a corresponding (simplified) C program produced from the second derivation.
\end{example}

\subsubsection{Behavior Blocks}
\label{sss_comportamento}

The dynamic behavior of \benchgen{} programs emerges from interactions with data structures.
All data structures within a program must share the same type.
Currently, \benchgen{} supports arrays of integers and sorted linked lists of integers.
Additionally, the C code generator also supports any of the twelve data structures available in the GNOME Library.
Adding support for new containers involves extending four C++ classes.
These classes define the behavior of four constructs in the L-grammar:

\begin{itemize}
\item \textbf{new:} Creates and initializes a data structure.
\item \textbf{insert:} Adds an element to a data structure in scope.
\item \textbf{remove:} Deletes an element from a data structure in scope.
\item \textbf{contains:} Checks whether an element exists in a data structure in scope.
\end{itemize}
Notice that these four behavior blocks can be customized by \benchgen{} users in any way.
For instance, \benchgen{} provides a ``{\it scalar mode}'' to generate programs that manipulate integer variables.
In this case, we have configured \texttt{insert} to increment variables; \texttt{remove} to decrement them; \texttt{new} to declare and initialize them with zero; and \texttt{contains} to test if they are zero.
As a more concrete example, below we show how three such operations may be defined for programs working with arrays.

\begin{figure}[t]
\centering
\includegraphics[width=0.7\columnwidth]{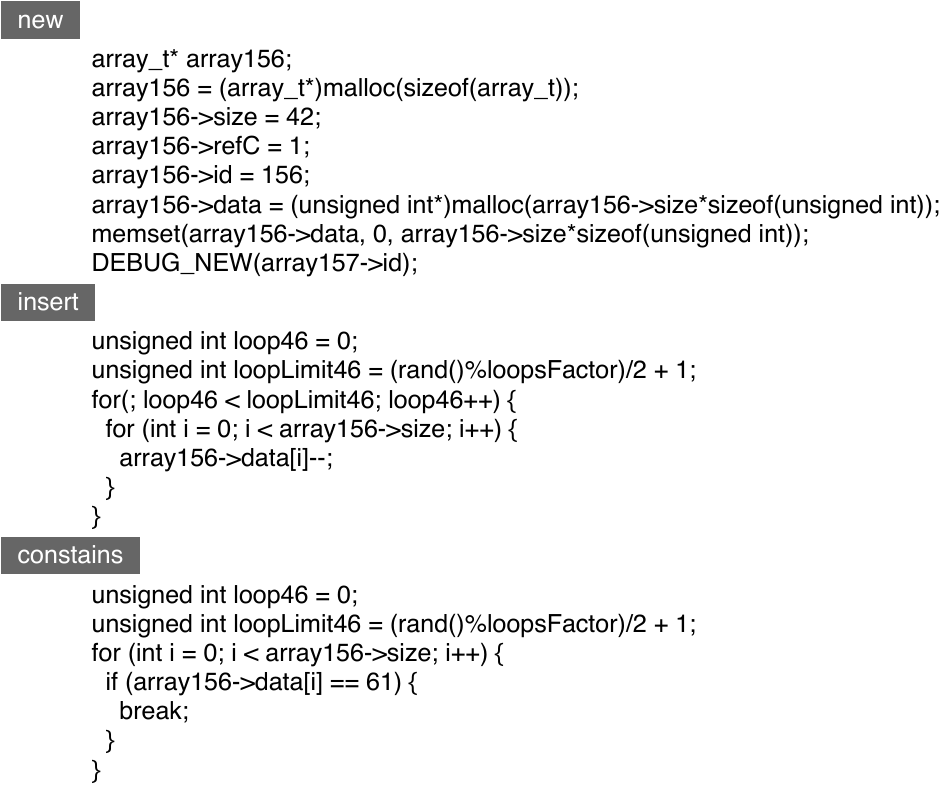}
\caption{Examples of block definitions for \texttt{new}, \texttt{insert}, and \texttt{contains}.}
\label{fig_behavioralOperations}
\end{figure}

\begin{example}
\label{ex_behavior}
Figure~\ref{fig_behavioralOperations} shows C code generated to manipulate arrays. Although a program may handle many arrays, the example focuses on a specific variable, \texttt{array156}. The operations \texttt{new()}, \texttt{insert()}, and \texttt{contains()} must be customized by the user to define the behavior of the generated code.
\end{example}

\subsection{Execution Flow}
\label{sub_controlFlow}

Programs generated by \benchgen{} are executable.
Their control flow is governed by a variable named \texttt{PATH}, of type \texttt{unsigned long}, which determines the outcome of conditional branches.
Algorithm BitPath in Figure~\ref{alg_bit_path} correlates the outcome of branches with this \texttt{PATH} variable.
The algorithm use a counter stack to ensure each path is uniquely identifiable via the \texttt{PATH} variable.
It assigns a unique bitmask to each branch condition by:
\begin{enumerate}
\item \textbf{Tracking Nesting Depth}: Using a stack to manage counters for each level of nested branches.
\item \textbf{Resolving Joins}: Propagating the "maximum counter" upward when branches merge, ensuring no bitmask collisions.
\end{enumerate}
The way Algorithm~\ref{alg_bit_path} determines how control-flow is taken lets us use \benchgen{} to test the effectiveness of profile-guided optimizations, as Section~\ref{sub_profile} will demonstrate.
This mechanism is clarified in Example~\ref{ex_controlFlow}.

\begin{algorithm}[ht]
\caption{BitPath assignment of unique bitmasks to branch conditions}
\label{alg_bit_path}
\KwIn{A function $f$}
\KwOut{A mapping of each branch condition in $f$ to a unique bitmask}

\textbf{Initialize the Stack}: push a single counter $Counter_0 = 1$, representing the least significant bit (LSB)\;

\BlankLine
\textbf{Assign Masks to Branches}:\;
\Indp
  \textbf{Same Nesting Level}: use the current top-of-stack counter to generate the bitmask, e.g., \texttt{PATH \& (1 << (counter - 1))}\;
  \textbf{Nested Branch}: push a new counter, incremented by 1 from the parent counter, ensuring fresh bits for deeper nesting\;
\Indm

\BlankLine
\textbf{Handle Join Points (after \texttt{if-then-else})}:\;
\Indp
  Pop the current counter from the stack\;
  Update the parent counter (now on top of the stack) to the maximum of:\;
  \Indp
    (a) the parent’s original value; \\
    (b) the popped (child) counter\;
  \Indm
This ensures that parent branches account for the deepest nesting level beneath them\;
\Indm

\BlankLine
\textbf{Independent Branches}: after resolving a join, subsequent branches at the same level reuse the updated parent counter\;

\end{algorithm}

\begin{figure}[t]
\centering
\includegraphics[width=0.8\columnwidth]{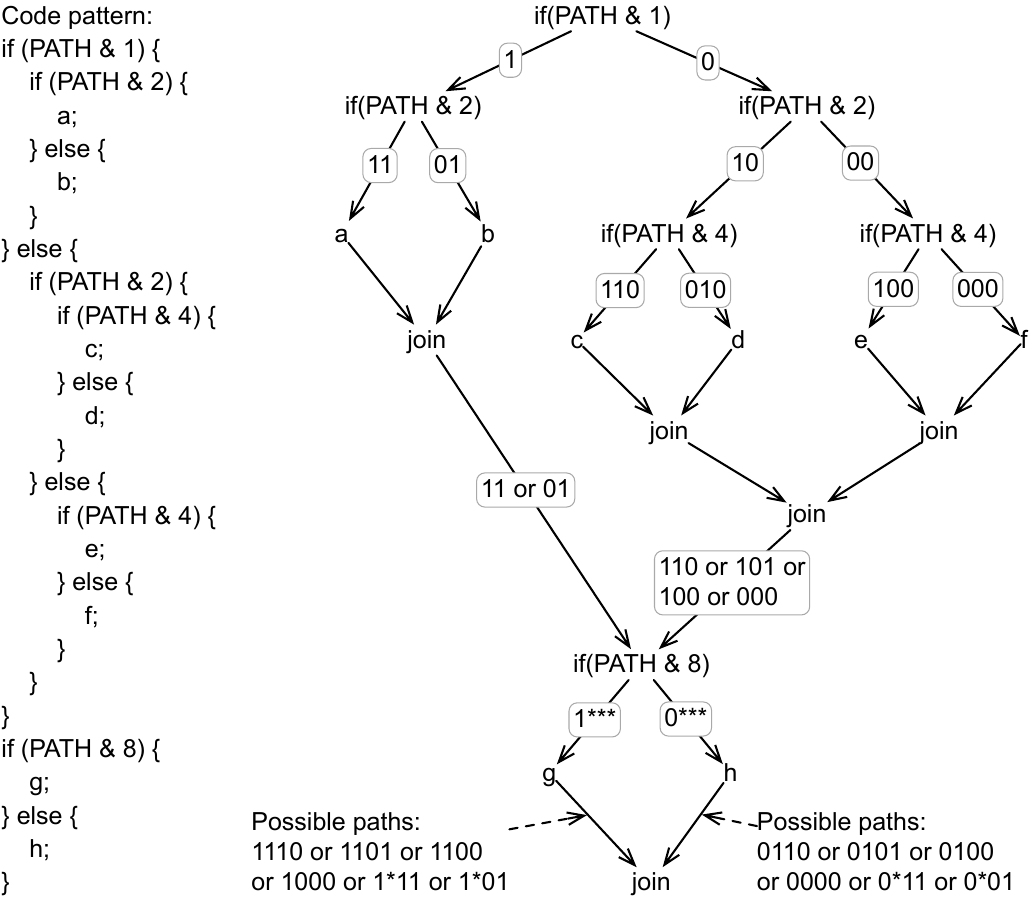}
\caption{The control flow of a synthetic program is determined by the \texttt{path} parameter.}
\label{fig_pathNumbering}
\end{figure}

\begin{example}
\label{ex_controlFlow}
Figure~\ref{fig_pathNumbering} depicts the control flow graph of a program with the \texttt{PATH} variable combined with masks created by the Algorithm~\ref{alg_bit_path}.
Notice that the counter used as the bitmask is updated due to nesting or sequencing.
Nesting causes the increment of the counter on the top of the counter stack.
Whenever we analyze a branch, its counter will be the maximum of all the possible control-flows that reacher it.
\end{example}

\subsection{Function Calls}
\label{sub_call}

\benchgen{} supports function calls through the \texttt{CALL} clause. When an L-string includes a construction of the form \texttt{CALL}($e$), the entire substring $e$ is extracted and defined as a separate function. The functions generated from a given L-specification can be either grouped into a single file or distributed across multiple files, depending on a configuration parameter in \benchgen{}.
Example~\ref{ex_functionCall} illustrates how \benchgen{} generates interprocedural code.

\begin{figure}[t]
\centering
\includegraphics[width=1\columnwidth]{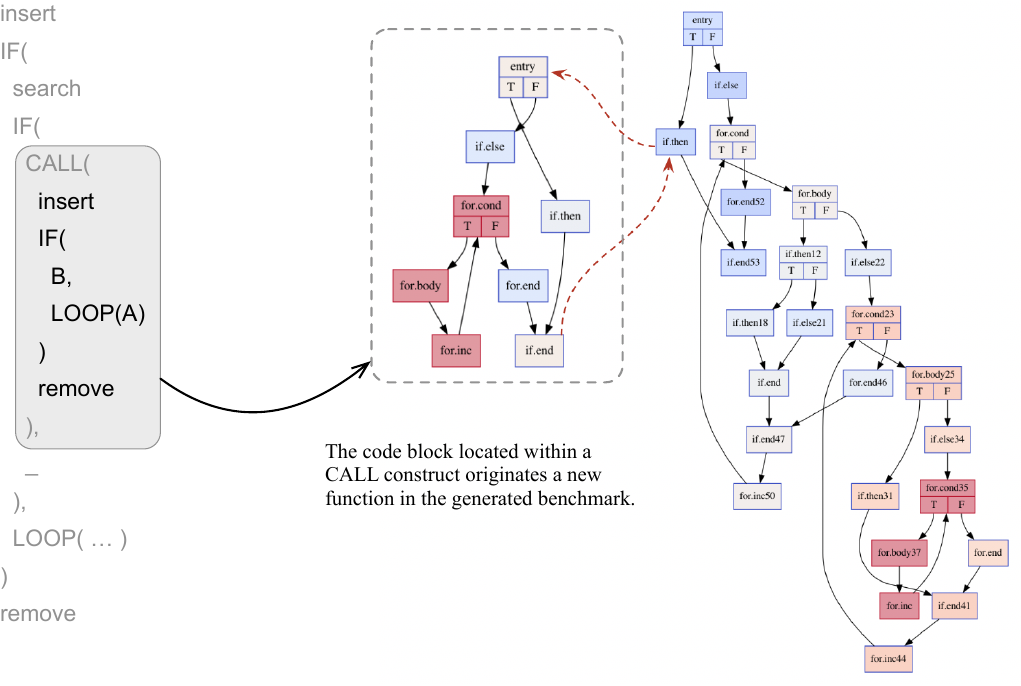}
\caption{Control flow of a program containing a \texttt{CALL} block.}
\label{fig_functionCall}
\end{figure}

\begin{example}
\label{ex_functionCall}
Figure~\ref{fig_functionCall} depicts the control flow produced by a \texttt{CALL} block. The enclosed string gives rise to a new function, which becomes part of the synthesized program. This new function is invoked at the point in the L-string where the \texttt{CALL} clause appears.
\end{example}

All occurrences of $\mathtt{CALL}(b)$ with identical strings $b$ result in calls to the same function. To avoid redundancy, \benchgen{} maintains a table mapping strings to functions, ensuring that identical strings refer to the same function instance. Example~\ref{ex_multipleFunctions} shows how the same function can be called multiple times.
Notice that \benchgen{} does not support the creation of recursive function calls.
If a function \texttt{f0} calls another function \texttt{f1}, then \texttt{f0} is produced by a string that is strictly larger than the string that generates \texttt{f1}.

\begin{figure}[t]
\centering
\includegraphics[width=0.7\columnwidth]{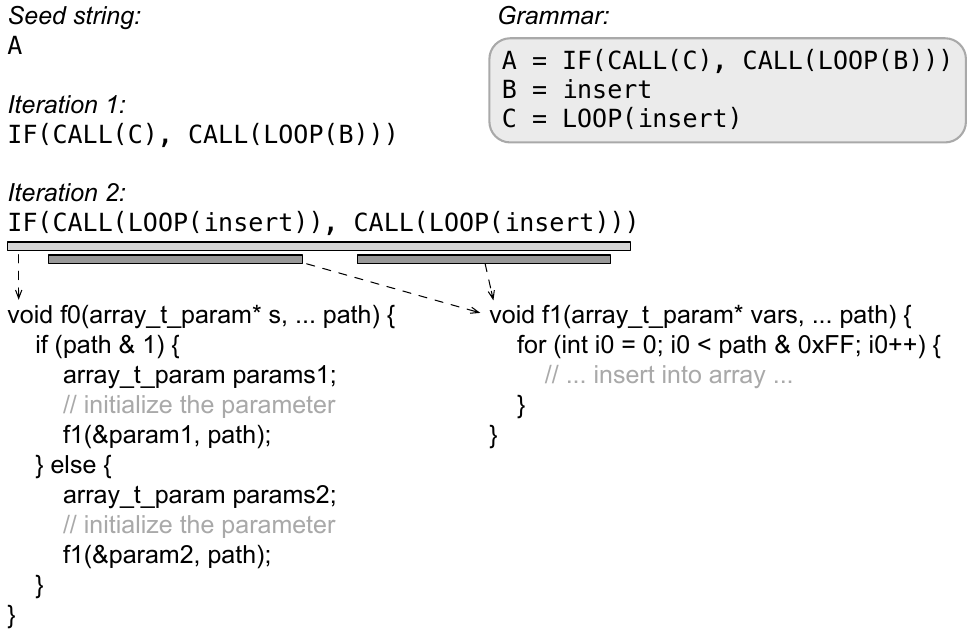}
\caption{An L-System that would generate multiple calls to the same function.}
\label{fig_multipleFunctions}
\end{figure}

\begin{example}
\label{ex_multipleFunctions}
Figure~\ref{fig_multipleFunctions} shows an L-System that generates a program with two functions, \texttt{f0} and \texttt{f1}.
The latter is called twice, because it is produced by a string that appears two times in a given iteration of the expansion process.
\end{example}

\paragraph{Parameter Passing. }
Functions generated by \benchgen{} receive two parameters:
\begin{itemize}
\item \textbf{Data:} an array of data structures that enables sharing between caller and callee functions.
\item \textbf{Path:} a control variable that governs execution flow, as described in Example~\ref{ex_controlFlow}.
\end{itemize}
The \textbf{Data} array is populated using a reaching definitions analysis, which determines which variables in the caller function are available to be passed as arguments at the call site. Example~\ref{ex_param} illustrates this mechanism.

\begin{figure}[t]
\centering
\includegraphics[width=0.7\columnwidth]{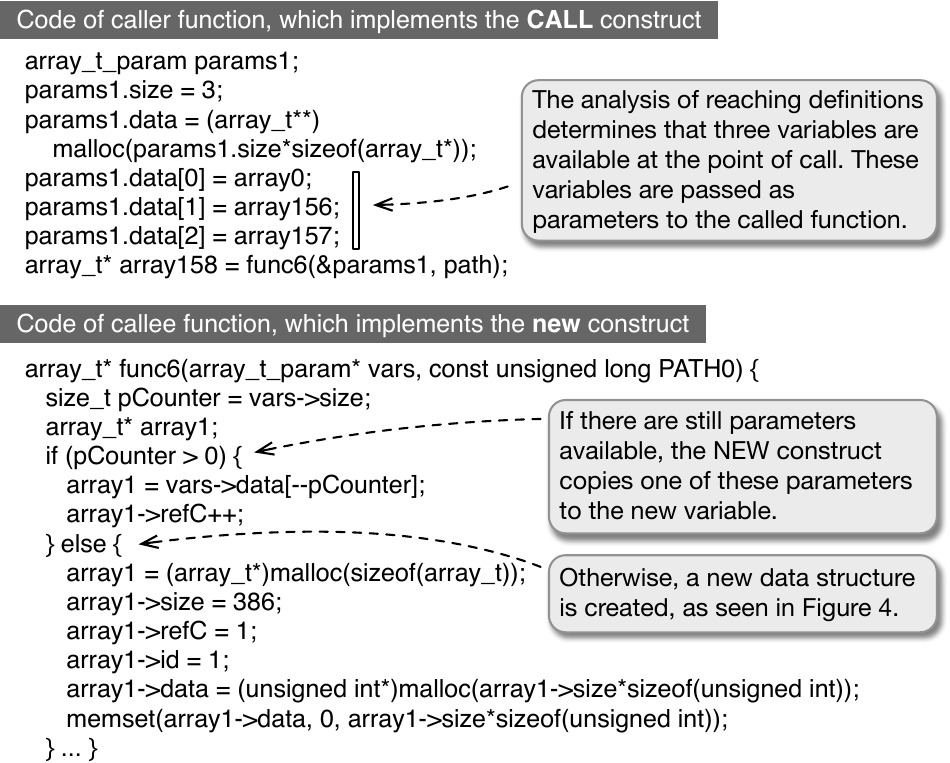}
\caption{Parameter passing in programs created by \benchgen{}.}
\label{fig_param_passing}
\end{figure}

\begin{example}
\label{ex_param}
Figure~\ref{fig_param_passing} illustrates parameter passing in a program generated by \benchgen{}. At the function call site, a reaching definitions analysis identifies three available variables: \texttt{array1}, \texttt{array156}, and \texttt{array157}. Pointers to these variables are inserted into the \texttt{param.data} structure, which is passed to function \texttt{func0}. Inside the calleer, the passed variables are copied into new ones using the \texttt{new} construct. Each variable is copied exactly once. If no additional parameters are available for copying, subsequent \texttt{new} operations will create fresh variables, as seen in Example~\ref{ex_controlFlow}.
\end{example}

\subsection{Memory Management}
\label{sub_memoryManagement}

Programs generated by \benchgen{} do not suffer from memory leaks, despite frequently relying on heap allocation.
To prevent leaks, \benchgen{} employs a reference-counting garbage collector. This approach tracks how many references (pointers) exist to each dynamically allocated object.
When a reference is created, the count is incremented; when it is removed, the count is decremented. Once the count reaches zero, the object is no longer reachable and can be safely deallocated. To support this mechanism, each structure created by \benchgen{} includes an additional field, \texttt{refC}, which stores the current reference count.
Example~\ref{ex_refCounter} shows how reference counting is used in \benchgen{} to prevent memory leaks from happening.

\begin{figure}[t]
\centering
\includegraphics[width=0.7\columnwidth]{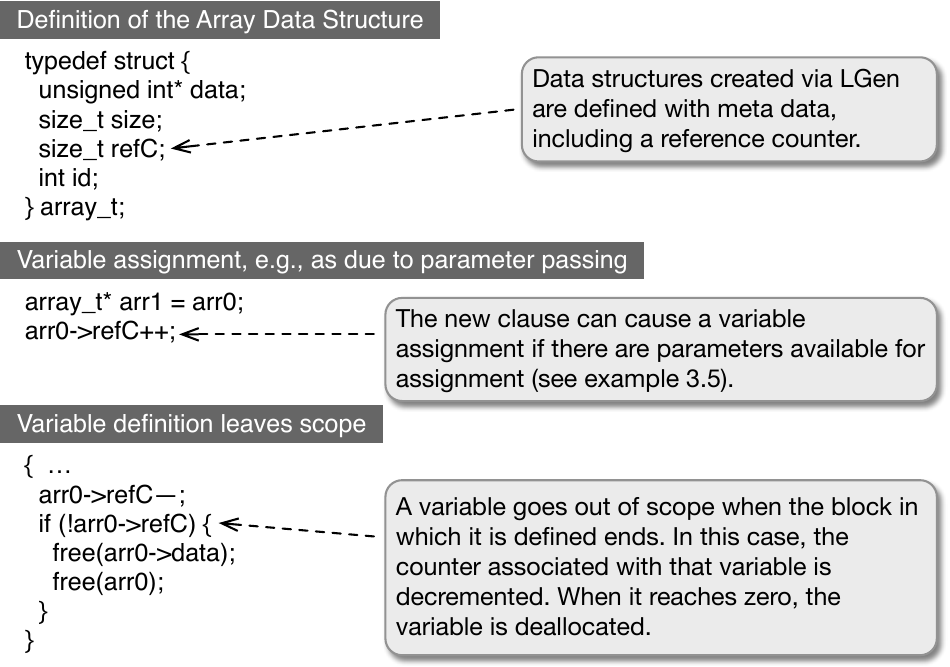}
\caption{Reference counter implementation.}
\label{fig_refCounter}
\end{figure}

\begin{example}
\label{ex_refCounter}
Figure~\ref{fig_refCounter} illustrates how \benchgen{} uses reference counting to manage memory. In this example, two variables, \texttt{x} and \texttt{y}, initially point to two separate heap-allocated structures. Each of these structures contains a \texttt{refC} field that holds the number of active references to the object.
When the assignment \texttt{x := y} occurs, the reference count of the structure originally pointed to by \texttt{x} is decremented, as \texttt{x} no longer refers to it. If this decrement causes \texttt{refC} to reach zero, the structure is automatically deallocated. Meanwhile, the reference count of the structure pointed to by \texttt{y} is incremented to account for the new reference from \texttt{x}. After the assignment, both \texttt{x} and \texttt{y} point to the same object, whose \texttt{refC} now reflects two active references.
This mechanism ensures that heap-allocated memory is reclaimed as soon as it is no longer reachable, preventing memory leaks in programs generated by \benchgen{}.
\end{example}

\subsection{Multilanguage Support}
\label{sub_multilang}

BenchGen is currently available in a beta version that supports the creation of benchmarks for multiple programming languages.
Some of these languages are evaluated in Section~\ref{sub_multilang_comparison}.
The benchmark generator itself is implemented in C++.
To add support for a new language, users must implement new C++ classes that define the semantics of the structure and behavior blocks described in Section~\ref{sub_buildingBlocks}.
This process also requires modifying the templates that generate code for the structure blocks (\texttt{IF}, \texttt{LOOP}, and \texttt{CALL}) and the behavior blocks (\texttt{insert}, \texttt{remove}, \texttt{contains}, and \texttt{new}).

Extending \benchgen{} therefore involves recompilation and is not yet fully automated.
In addition to implementing new C++ classes, users must register the language module in the main \benchgen{} driver, which requires changing one line of code in the driver itself.
In future work, we aim to fully automate this process and decouple the tool from the specific languages it supports.
To this end, we are considering extending \benchgen{} with a domain-specific language that would allow users to specify new code-generation directives declaratively.

\section{Experimental Evaluation}
\label{sec_eval}

The goal of this section is to demonstrate how \benchgen{} can be used in practice.  
To this end, we explore its usage in the following {\it Case Studies}:
\begin{itemize}
\item \textbf{CS1}: Comparison between \texttt{gcc} and \texttt{clang} in terms of execution speed, binary size, and compilation time.
\item \textbf{CS2}: Comparison of the C and C++ compiler backends available in the \textsc{Gnu} Compiler Collection.
\item \textbf{CS3}: Analysis of the asymptotic behavior of different optimization levels in \texttt{clang} and \texttt{gcc}.
\item \textbf{CS4}: Examination of the evolution of \texttt{gcc} from version 5 to 14, focusing on execution speed, binary size, and compilation time.
\item \textbf{CS5}: Evaluation of the impact of profile-guided optimizations in \texttt{clang}.
\item \textbf{CS6}: Comparison of different data structures in the \textsc{Gnome Library} (\textsc{GLib}) with respect to insertion, search, and deletion times.
\item \textbf{CS7}: Comparison between \benchgen{} and \textsc{CSmith}, a compiler fuzzer.
\end{itemize}

\textbf{Experimental Setup:}  
All experiments were conducted on an Intel(R) Xeon(R) CPU E5-2680 v2 running at 2.80\,GHz, with Linux Ubuntu 5.15.0-139-generic.  
The specific compiler and library versions used are detailed in each case study.

\textbf{Checking for Undefined Behavior:}
We analyzed the C programs evaluated in Section~\ref{sub_gcc_vs_clang} using four tools capable of detecting undefined behavior:
\textsc{UBSan} and \textsc{ASan} from \texttt{clang}~\cite{Serebryany12}, Valgrind~\cite{Nethercote07}, the EVA plugin of Frama-C~\cite{Baudin21}, and KCC~\cite{Ellison12}.
In addition, we verified that every program evaluated in Section~\ref{sub_multilang_comparison}, when executed in debug mode, produces the same trace\footnote{\benchgen{} provides a debug mode in which each program logs every behavioral operation defined in Section~\ref{sss_comportamento}.}.
While these measures do not formally prove that \benchgen{} always generates sound benchmarks, they provide strong evidence that this is highly probable.


\subsection{CS1: \texttt{gcc} vs \texttt{clang}}
\label{sub_gcc_vs_clang}

The GNU C Compiler (\texttt{gcc}) and LLVM's \texttt{clang} are the two most widely used C compilers, and comparisons between them are common in the literature.
This case study contributes one more data point to this ongoing discussion by evaluating both compilers across three performance metrics:
\begin{itemize}
\item Execution time of the compiled program.
\item Compilation time required to build the program.
\item Binary size of the compiled program.
\end{itemize}
The comparison covers six optimization levels for \texttt{gcc} 14.2: \texttt{-O0}, \texttt{-O1}, \texttt{-O2}, \texttt{-O3}, \texttt{-Os}, and \texttt{-Ofast}; and seven for \texttt{clang} 21.0: \texttt{-O0}, \texttt{-O1}, \texttt{-O2}, \texttt{-O3}, \texttt{-Oz}, \texttt{-Os}, and \texttt{-Ofast}.
Binary sizes were collected using the Linux \texttt{size} command, considering only the \texttt{text} section, which represents the size of the executable code in bytes.
Compilation and execution times were measured with \texttt{hyperfine}, configured with three warm-up runs and at least ten benchmark runs.
To compare the compilers, we have used four different L-Systems to generate four programs.
We chose generations that ensured a mix of different execution times; hence, this custom benchmark collection contains programs that run within 10, 50 and 200 seconds once compiled with \texttt{gcc -O3}.

\begin{figure}[t]
\centering
\includegraphics[width=1\textwidth]{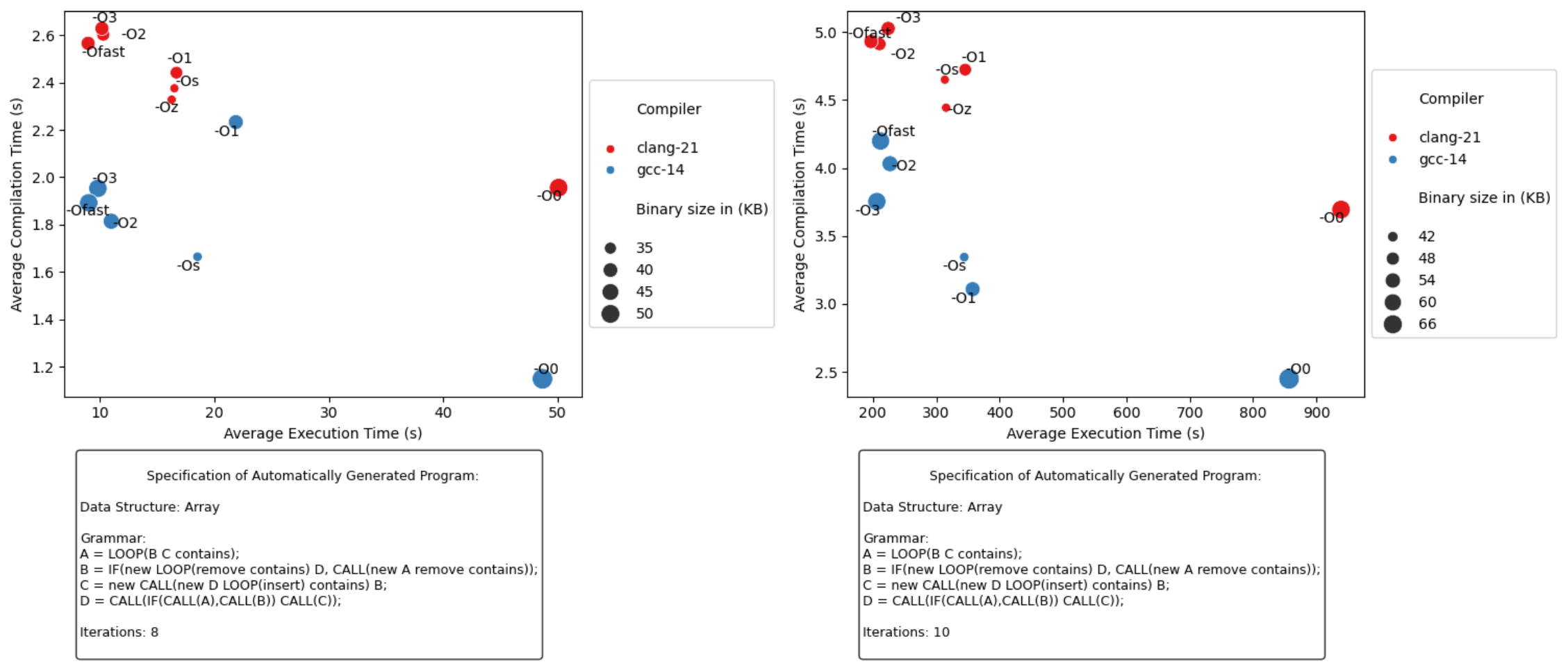}
\caption{Comparison of \texttt{gcc} 14.2 and \texttt{clang} 21.0 across optimization levels (Part 1).}
\label{fig_compilers_experiment_1}
\end{figure}

\begin{figure}[t]
\centering
\includegraphics[width=1\textwidth]{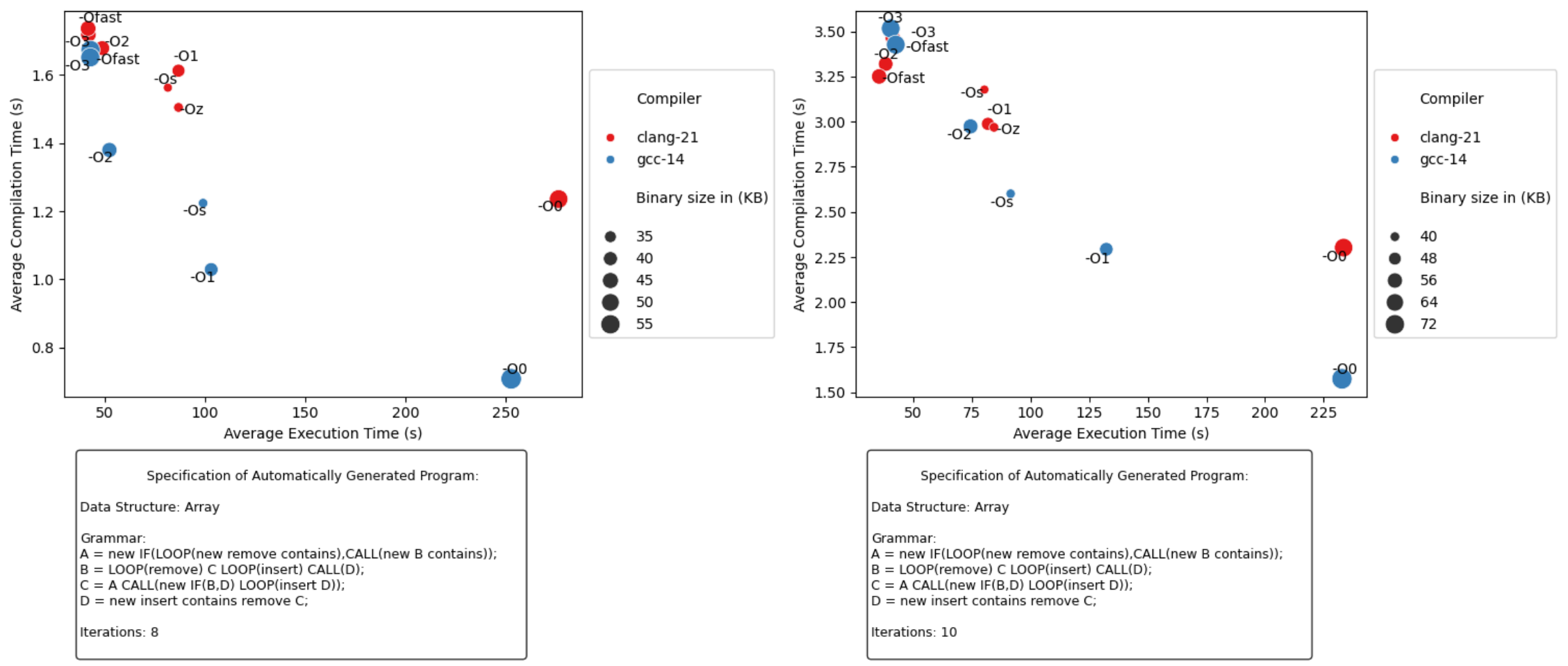}
\caption{Comparison of \texttt{gcc} 14.2 and \texttt{clang} 21.0 across optimization levels (Part 2).}
\label{fig_compilers_experiment_2}
\end{figure}

\paragraph{Discussion. }
Figures~\ref{fig_compilers_experiment_1} and~\ref{fig_compilers_experiment_2} summarize the results of this comparison, showing, under each plot, the L-System and the generation that produced those numbers.
The scatter plots show compilation time against execution time, with point size proportional to binary size.
We observe that \texttt{gcc} and \texttt{clang} tend to produce programs with similar execution times at the lowest and highest optimization levels; however, \texttt{gcc}'s compilation times are generally lower.

A clear trend emerges for \texttt{gcc}: it often exhibits a smooth trade-off between compilation time and execution time, forming a Pareto-like curve where longer compilation typically results in faster execution.
This pattern is less evident for \texttt{clang}.
Instead, \texttt{clang} shows a marked distinction between the lowest optimization levels (\texttt{-O0}, \texttt{-O1}) and the higher ones, while the most aggressive optimizations (\texttt{-O2}, \texttt{-O3}, and \texttt{-Ofast}) yield very similar results.
This observation echoes findings by Curtsinger and Berger in their work on Coz~\cite{Curtsinger15}, who reported no statistically significant difference between \texttt{-O2} and \texttt{-O3} in \texttt{clang}\footnote{See \url{https://youtu.be/r-TLSBdHe1A?t=1438} at 23:58}.
Regarding binary size, both compilers generate smaller executables at \texttt{-Os}, with \texttt{clang} holding a slight edge (2--3\%) at the \texttt{-Oz} level.
Nevertheless, both can reduce code size by nearly 50\% when moving from \texttt{-O3} to \texttt{-Os}.

\subsection{CS2: Comparison between Different Programming Languages}
\label{sub_multilang_comparison}

\begin{figure}[t]
\centering
\includegraphics[width=0.8\columnwidth]{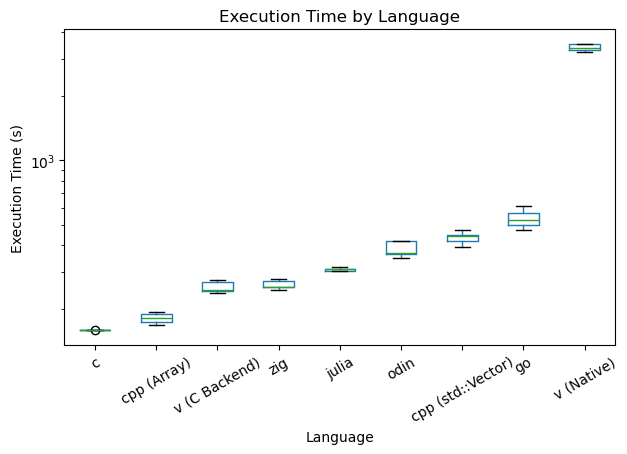}
\caption{Execution time of benchmarks produced by \benchgen{} in different programming languages.}
\label{fig_multilangComparison}
\end{figure}

As explained in Section~\ref{sub_multilang}, \benchgen{} can be customized to synthesize benchmarks in multiple programming languages.
This section leverages its multi-language support to compare C, C++, Julia, Go, Zig, V and Odin.
Support for these languages is currently available in \benchgen{}'s official repository.
To perform this comparison, we used the first grammar shown in Figure~\ref{fig_compilers_experiment_1} to produce eleven generations of a program, and report five runs of the 11th.
Programs in each language were executed with the following tools:
\begin{description}
\item[C:] \texttt{gcc} 13.3.0
\item[C++:] \texttt{g++} 13.3.0
\item[Go:] \texttt{go} 1.25.0
\item[Julia:] \texttt{julia} 1.11.6
\item[Odin:] \texttt{odin} dev-2015-11
\item[V:] \texttt{vlang} 0.4.12
\item[Zig:] \texttt{zig} 0.16.0
\end{description}
Note that Julia runs in interpreted mode with just-in-time compilation, whereas the other languages are compiled ahead of time.
V has two execution modes: native and C-backend.
The former directly compiles V to machine code, whereas the latter first translates V to C, and then uses \texttt{clang-22} to produce binary code out of this C source.
In this case, we use the optimization level \texttt{-O3}, as we do in C and C++.
We include two versions of C++, one using plain arrays and another using \texttt{std:vector}, just to give the reader some perspective on the relative running times that Figure~\ref{fig_multilangComparison} presents.
This evaluation considers only execution time, excluding compilation time.
Also, each language produces the same number of source files, except for C and C++, which generate one extra header file.
For example, in the 11th generation, \benchgen{} produces 52 files for Julia and Go, and 53 for C and C++.

\paragraph{Discussion. }
Figure~\ref{fig_multilangComparison} compares execution time across languages.
The results largely align with expectations: they highlight the performance differences between low-level, statically compiled languages (C and C++) and higher-level languages with different compilation models (Go and Julia).
Although we omit the running time of programs in generations 1 to 10, we emphasize that all languages exhibit an exponential increase in execution time as program depth grows.
This behavior is inherent to the fractal-based benchmark generator.
Beyond this general trend, we note several specific findings:
\begin{itemize}
\item \textbf{C vs. C++ with arrays.} Performance is nearly identical: there is no statistically significant difference at the 95\% confidence level when comparing programs of the same generation. This is because both are compiled with the same compiler infrastructure (\texttt{gcc} 13.3.0), which produces highly similar assembly code.

\item \textbf{Julia.} Julia shows excellent performance, particularly for larger programs---a behavior already observed in previous work~\cite{Bezanson18}. Its just-in-time (JIT) compiler, built on LLVM, can apply aggressive runtime optimizations such as loop unrolling and SIMD vectorization by specializing code based on runtime types and execution patterns. As program complexity increases in later generations, these optimizations give Julia a significant advantage, allowing it to outperform C++ with vectors.

\item \textbf{C++ with vectors.} Programs using \texttt{std::vector} are slower than those using raw arrays. This overhead arises from additional metadata (size, capacity) and the cost of reallocation when capacity is exceeded, which requires copying elements to new memory. In benchmarks where data grows quickly, this reallocation cost is significant. Indeed, from the 8th to the 9th generation, Julia’s JIT optimizations allow it to surpass C++ with vectors in performance.

\item \textbf{Go.} Go is consistently the slowest language. This is unsurprising, and follows trends already reported and discussed in open forums and technical books~\cite{Strecansky20}. While Go is compiled, its runtime includes garbage collection and applies less aggressive optimizations, which introduces overhead not present in C, C++.

\item \textbf{V.} There is a significant difference between V's native compiler and the V-to-C translator.
The former takes, on average, about 3,200 seconds to run programs on the 11th generation, whereas the latter takes 270 seconds.
This difference outlines the impact of compiler optimizations, which are mostly absent on the native code generator, and used abundantly in the C backend.
\end{itemize}
The trends observed in Figure~\ref{fig_multilangComparison} are consistent with those documented on the well-known ``Benchmark Game'' website\footnote{Available at \url{https://benchmarksgame-team.pages.debian.net/benchmarksgame/} as of September 20th.}. A key distinction, however, lies in the implementation strategy. While the Benchmark Game compares hand-optimized solutions that may differ syntactically and semantically, \benchgen{} produces structurally similar code across languages. This methodology offers a more precise comparison of compiler performance by eliminating variability introduced by differing algorithmic implementations.

\subsection{CS3: Asymptotic Behavior}
\label{sub_asymptotic}

The ability to gradually vary the size of programs generated by \benchgen{} makes it a suitable tool for conducting empirical asymptotic analyses of language processing systems.
In this section, we illustrate this capability by empirically evaluating the asymptotic complexity of different phases of \texttt{clang} (front end, middle end, and back end), as well as the full compilation pipeline of \texttt{gcc}.
For this purpose, we employ a standard L-System to produce eight generations of a program (from the 5th to the 12th) and feed these programs to the compilers.

\begin{figure}[t]
\centering
\includegraphics[width=1\columnwidth]{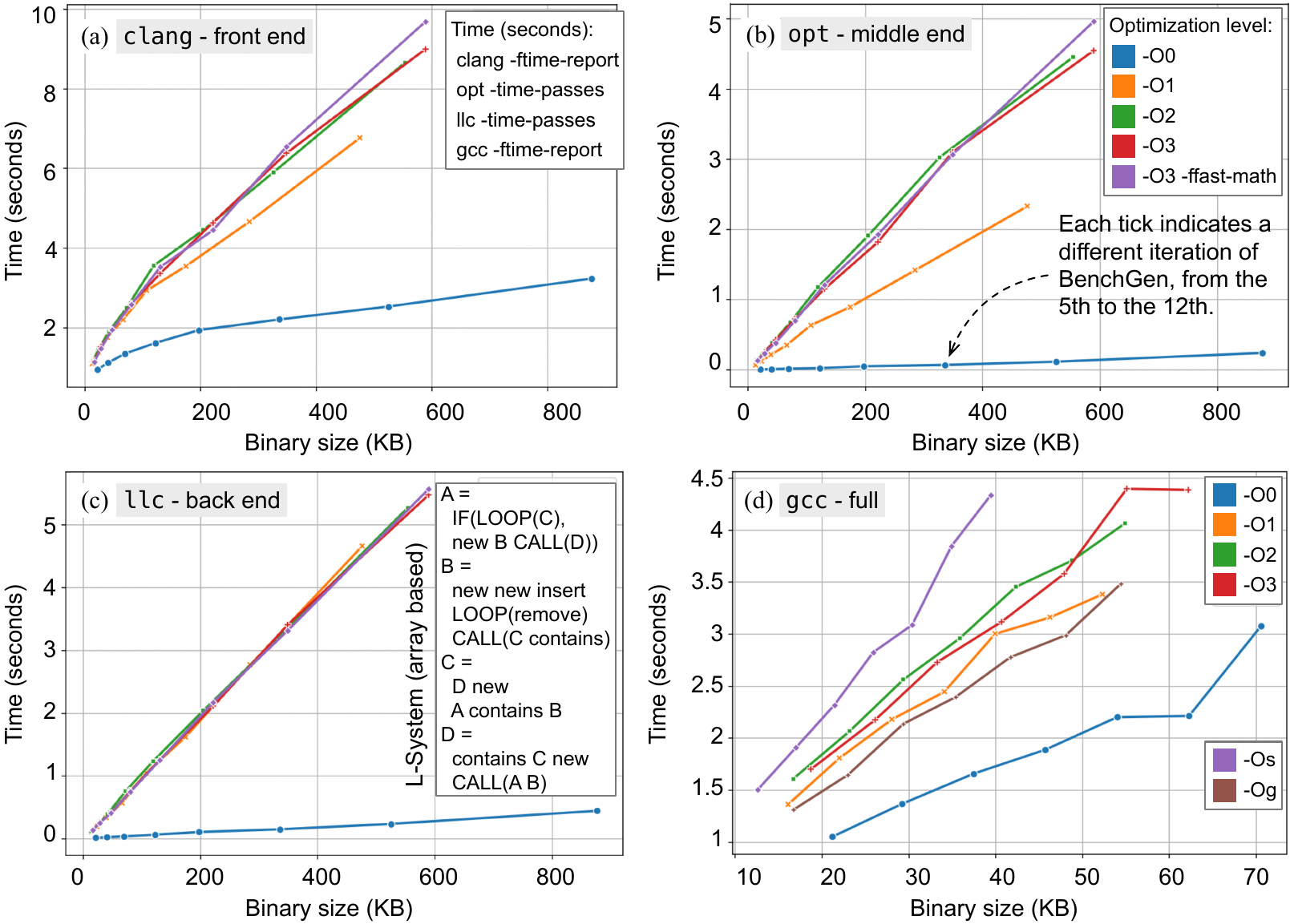}
\caption{Empirical evaluation of the asymptotic behavior of different compilation phases of \texttt{clang} and of the full compilation pipeline of \texttt{gcc}.}
\label{fig_clangAsymptotic}
\end{figure}

\paragraph{Discussion. }
Figure~\ref{fig_clangAsymptotic} relates the running time of the different compilation phases to the size of the binary processed at each generation of the target program.
This analysis yields several insights, discussed below:
\begin{itemize}
\item All analyzed tools exhibit linear behavior for large programs, with very strong correlations (above 0.9) across different compilers and optimization levels.
This applies to both \texttt{clang} and \texttt{gcc}.
However, for \texttt{clang}'s front end, linear growth only becomes evident in later generations (iterations 9--12), due to the heavy startup costs of the parser and IR generator.

\item There is little statistical difference among the higher optimization levels of \texttt{clang} (\texttt{-O2}, \texttt{-O3}, and \texttt{-Ofast}).
This observation corroborates the findings reported in Section~\ref{sub_gcc_vs_clang}.
It appears valid, at least in this experiment, across all three phases of \texttt{clang}'s compilation pipeline, and is especially pronounced in \texttt{llc}, the machine code generator.

\item Although \texttt{gcc} also shows strong linear behavior at all optimization levels, the constant factors vary considerably.
The size-optimization level (\texttt{-Os}) exhibits the steepest growth, which is expected since \texttt{gcc -Os} produces the smallest binaries.
\end{itemize}
Overall, these results suggest that, at least when compiling \benchgen{} programs, neither \texttt{clang} nor \texttt{gcc} exhibit asymptotic performance bugs of the kind reported in previous work~\cite{Olivo15,Han16}.

\subsection{CS4: The Evolution of \texttt{gcc}}
\label{sub_evolution}

GNU \texttt{gcc}, currently at version 14, is one of the oldest and most widely used C compilers. Over the years, it has undergone many changes, including support for additional languages and continuous performance improvements. This case study compares different versions of \texttt{gcc} across multiple optimization levels using three metrics:
\begin{itemize}
\item Execution time of the compiled program (measured with \texttt{hyperfine}).
\item Compilation time required to build the program (also measured with \texttt{hyperfine}).
\item Binary size of the \texttt{.text} segment of the executable (collected with the Linux \texttt{size} utility).
\end{itemize}
For this comparison, we used \benchgen{} with the first L-System shown in Figure~\ref{fig_compilers_experiment_1}. The program generated at the 10th iteration of this grammar was compiled with six versions of \texttt{gcc}.

\begin{figure}[t]
\centering
\includegraphics[width=1\columnwidth]{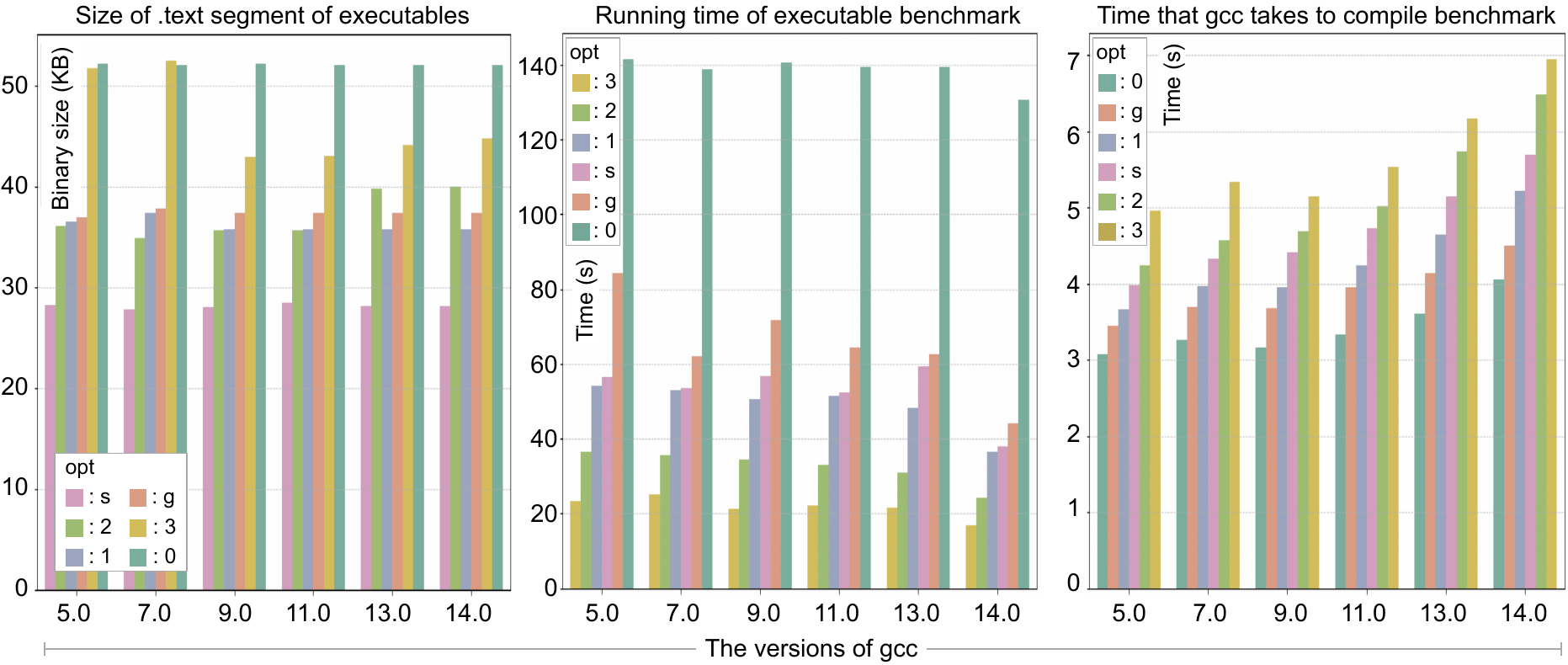}
\caption{Evolution of \texttt{gcc} from version 5 to 14 across six optimization levels. Results are shown for (left) binary size of the \texttt{.text} segment, (center) execution time of the benchmark, and (right) compilation time.}
\label{fig_gcc_evolution}
\end{figure}

\paragraph{Discussion. }
Figure~\ref{fig_gcc_evolution} summarizes results for six optimization levels: \texttt{-O0}, \texttt{-Og}, \texttt{-O1}, \texttt{-Os}, \texttt{-O2}, and \texttt{-O3}. Each group of bars corresponds to a compiler version: \texttt{gcc5}, \texttt{gcc7}, \texttt{gcc9}, \texttt{gcc11}, \texttt{gcc13}, and \texttt{gcc14}.
Below we discuss some of the insights of this study:

\begin{description}
\item[Compilation time:] We observe a steady increase in compilation time with each new version of \texttt{gcc}, particularly at higher optimization levels such as \texttt{-O3}. Comparing versions 5 and 14, compilation time rose by about 42\% over nine years, with a 31\% increase even at \texttt{-O0}. This is unsurprising: new optimization passes have been added over time, which increases compile time in exchange for better code quality.

\item[Execution time:] Performance has improved at the higher optimization levels. Comparing \texttt{gcc5} to \texttt{gcc14}, execution time decreased by roughly 27\% at \texttt{-O3}, 32\% at \texttt{-O1}, and as much as 47\% at \texttt{-Og}. The gains at \texttt{-O0} were modest, about 7\%. These results indicate that newer versions of \texttt{gcc} generate more efficient code, especially at intermediate optimization levels.

\item[Binary size:] Binary size has remained stable across versions. At \texttt{-Os}, code size decreased only 0.34\% between \texttt{gcc5} and \texttt{gcc14}. At \texttt{-O3}, binary size initially increased from \texttt{gcc5} until \texttt{gcc7}, dropping noticeably in \texttt{gcc9}.
Past this point, we again observe a small increase until \texttt{gcc14}.
However, in this version code is still almost 10\% smaller than in \texttt{gcc5}.
\end{description}
Over the evolution from version 5 to 14, \texttt{gcc} shows a clear trade-off: longer compilation times in exchange for more efficient executables. While binary size has changed little, execution performance has improved noticeably, particularly at \texttt{-Og} and \texttt{-O3}. These results highlight an important trend: the \texttt{gcc} community has consistently prioritized code quality and runtime efficiency over compilation speed, accepting longer build times in return for measurable performance gains.

\subsection{CS5: Profile Guided Optimizations in \texttt{clang}}
\label{sub_profile}

Profile-Guided Optimization (PGO) is a compilation technique that leverages runtime information collected from representative executions of a program to improve the quality of the generated code.
By replacing purely static heuristics with empirical execution data, PGO enables the compiler to make more informed decisions regarding branch prediction, function inlining, loop unrolling, and code layout, thereby enhancing both performance and efficiency.
In typical C, C++ and Rust compilers, PGO is integrated into the standard optimization pipeline rather than constituting a separate optimization level. Specifically in \texttt{clang}, PGO serves as a refinement to existing optimization levels, most notably \texttt{-O2} and \texttt{-O3}.
When profile data is provided via the \texttt{-fprofile-use} flag, \texttt{clang} applies the chosen baseline optimizations but augments them with profile-driven insights, generating code that more accurately reflects the observed runtime behavior of the program.

This section illustrates how \benchgen{} can be used to evaluate the effectiveness of profile-guided optimizations (PGO).  
As described in Section~\ref{sub_controlFlow}, the control flow of a program generated by \benchgen{} is governed by the variable \texttt{PATH}.  
To assess the performance gains provided by PGO, we conduct the following experiment:  

\begin{enumerate}  
\item Generate four program variants (Generations 6, 7, 8, and 9) using the following grammar with array structures:
\begin{verbatim}  
A = new B B  
B = IF(LOOP(insert A contains), LOOP(insert A contains))  
\end{verbatim}  

\item Compile each program at the \texttt{-O2} optimization level with PGO enabled\footnote{We report results for \texttt{clang} at the \texttt{-O2} level because the speedups achieved with PGO are comparable to those observed at \texttt{-O3}, and consistently higher than those obtained at \texttt{-O1}.}.  

\item During the ``{\it Training Phase},'' collect profile information by running each program with $\mathtt{PATH} = 2^0$ (i.e., $\mathtt{PATH} = 1$).  

\item During the ``{\it Testing Phase},'' for $i = 1 \ \mathtt{to}\ 63$:  
\begin{enumerate}  
\item Execute each program with $\mathtt{PATH} = 2^i - 1$, e.g., $\mathtt{PATH} = 0b1$, $\mathtt{PATH} = 0b11$, etc.  
\end{enumerate}  
\end{enumerate}  
Let $t$ denote the execution time of the program compiled with \texttt{clang -O2} without profile data, and let $t_i$ denote the execution time with $\mathtt{PATH} = 2^i$.  
Then, the ratio $t/t_i$ quantifies the relative speedup due to profile-guided optimizations as the control-flow path diverges by $i$ bits.

\begin{figure}[t]
\centering
\includegraphics[width=1\columnwidth]{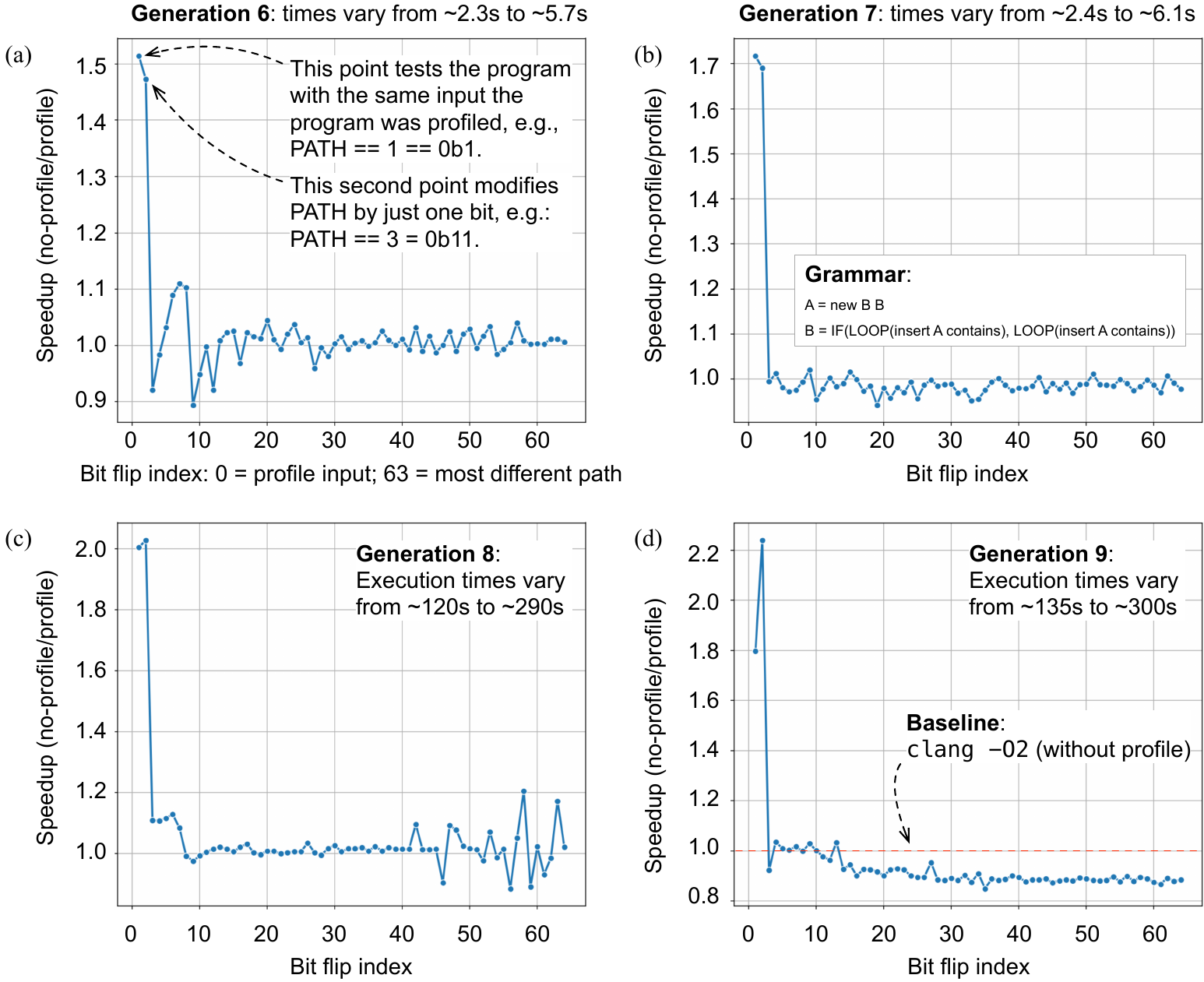}
\caption{The impact of profile guided optimizations on \texttt{clang -O2}.}
\label{fig_profileExperiment}
\end{figure}

\paragraph{Discussion. }
Figure~\ref{fig_profileExperiment} shows the data collected during this experiment.
The results in Figure~\ref{fig_profileExperiment} reveal two distinct behaviors of profile-guided optimizations.
In the expected case (Generations 6–8), the performance trend matches our intuition: when the execution path during testing closely resembles the path exercised during training, we observe significant speedups, often reaching 2x.
Similar results on real applications have been reported in previous work~\cite{Dehao16,Panchenko19,Moreira21}.

As the number of bit flips in \texttt{PATH} increases, however, the executed control flow diverges from the profiled run, and the benefit of PGO diminishes. Eventually, for highly divergent inputs, the profiled and non-profiled binaries converge to similar performance, as the compiler’s profile-driven decisions no longer apply.

\textbf{Profile Overfitting: }
The unexpected case arises in Generation~9.
Here, PGO yields a very large speedup when execution follows the training path (more than 2.2x).
Yet, as the path diverges, the optimized binary becomes slower than its non-profiled counterpart. This behavior suggests an instance of profile overfitting: because PGO biases optimizations toward the observed hot path, code along unprofiled paths may suffer from unfavorable decisions.

By reading \texttt{perf} counters, we observe that the program optimized with profile data experiences more cache references and cache misses than the program optimized without it, as the training and test path diverge.
The ratio of branch misses, in contrast, do not show more variance as the paths diverge.
A plausible explanation for this effect is the difference in the number of functions inlined with and without profiling data.
Clang/LLVM makes inlining decisions at the call-site level.
A function call along the hot path may be inlined aggressively, while the same call in a cold path may be left uninlined, leading to inconsistencies in call overhead, code size, and optimization opportunities.
Combined with profile-driven code layout and branch prediction, this selective optimization can make non-profiled paths slower than in a neutral, non-PGO binary.
Thus, Generation~9 highlights a limitation of PGO: while it can deliver substantial performance gains when training inputs are representative, it may degrade performance when profile data overfits to a narrow subset of execution paths.

2\subsection{CS6: Data Structures in \textsc{GLib}}
\label{sub_glib}

The GNOME Library (\textsc{GLib}) is a C library developed as part of the GNOME project that provides a wide range of data structure implementations, utility functions, and portability abstractions.
Among its generic containers are dynamic arrays (e.g., \textsc{GArray}, \textsc{GPtrArray}), byte arrays (\textsc{GByteArray}), singly linked lists (\textsc{GSList}), doubly linked lists (\textsc{GList}), double-ended queues (\textsc{GQueue}), hash tables (\textsc{GHashTable}), balanced binary trees (\textsc{GTree}), and others. 
This variety exists because no single container is optimal for all purposes: some favor fast insertion or deletion (especially at specific positions), others provide efficient random access, better memory locality, sorted traversal, or minimal overhead.
By offering multiple data structures, \textsc{GLib} allows programmers to select the container best suited to their performance, memory, and semantic requirements.

\benchgen{} enables the systematic evaluation of the dynamic behavior of these data structures.  
As discussed in Section~\ref{sss_comportamento}, the current implementation of the C code generator supports all twelve containers provided in the default distribution of the library\footnote{A detailed description of these containers is available at \url{https://docs.gtk.org/glib/data-structures.html}.}. 
This section demonstrates how such support can be used to compare the performance characteristics of different containers.

\paragraph{Experimental setup. }
We stress three operation classes: \texttt{insert}, \texttt{remove}, and \texttt{contains}. Programs are generated from a L-System that (i) quickly populates the container and (ii) expands a template where a placeholder is filled with a random sequence of operations:

\begin{verbatim}
A0 = A1 A1 A1 A1 A1 A1 A1 A1;
A1 = A2 A2 A2 A2 A2 A2 A2 A2;
A2 = A3 A3 A3 A3 A3 A3 A3 A3;
A3 = insert insert;
B = LOOP(CALL(B) C);
C = B <operations> B;
\end{verbatim}

The initialization gadget above yields $8^3\cdot 2 = 1024$ insertions; BenchGen must be run with depth $\ge 4$ for this expansion to produce operations. The \texttt{<operations>} placeholder is expanded into a random ordering of \texttt{insert}, \texttt{remove}, and \texttt{contains} tokens; we vary the count of each token over the set ${0,2,4,6,8,10}$, producing $6^3=216$ programs per container.

Each generated program is compiled twice: once with per-operation printing enabled (to count actual operation frequencies) and once with printing disabled (to measure runtime without I/O overhead). Binaries are benchmarked with \texttt{hyperfine}; due to the total runtime, we use a single warm-up run per measurement. Programs with similar numbers of a given operation are grouped, and we report the average running time per group.

\begin{figure}[t]
\centering
\includegraphics[width=1\columnwidth]{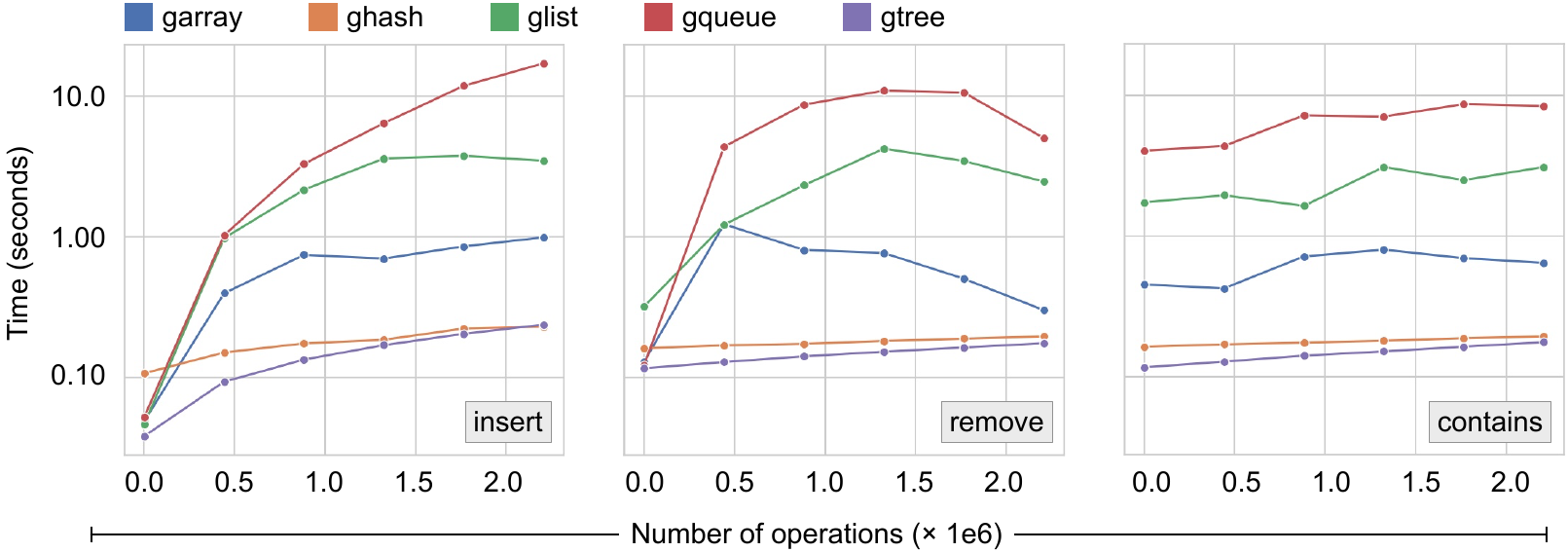}
\caption{Comparison of five data structures from \textsc{GLib}: \textsc{GHashTable}, \textsc{GTree}, \textsc{GArray}, \textsc{GQueue}, and \textsc{GList}.}
\label{fig_allDataStructures}
\end{figure}

\paragraph{Discussion. }
Figure~\ref{fig_allDataStructures} shows average running time as a function of the number of operations. Hash tables and balanced trees outperform lists, arrays, and queues as workloads grow. Among the latter, \textsc{GList} outperforms \textsc{GQueue}, which is expected because \textsc{GQueue} is implemented on top of \textsc{GList} and therefore inherits additional pointer-management overhead.

A noteworthy artifact in the results is that the average execution time decreases as the number of removals increases. This is not a correctness issue but a consequence of the grouping: removal-heavy programs tend to leave containers nearly empty, so subsequent operations touch far fewer elements and the per-group average runtime falls.

\begin{figure}[t]
\centering
\includegraphics[width=1\columnwidth]{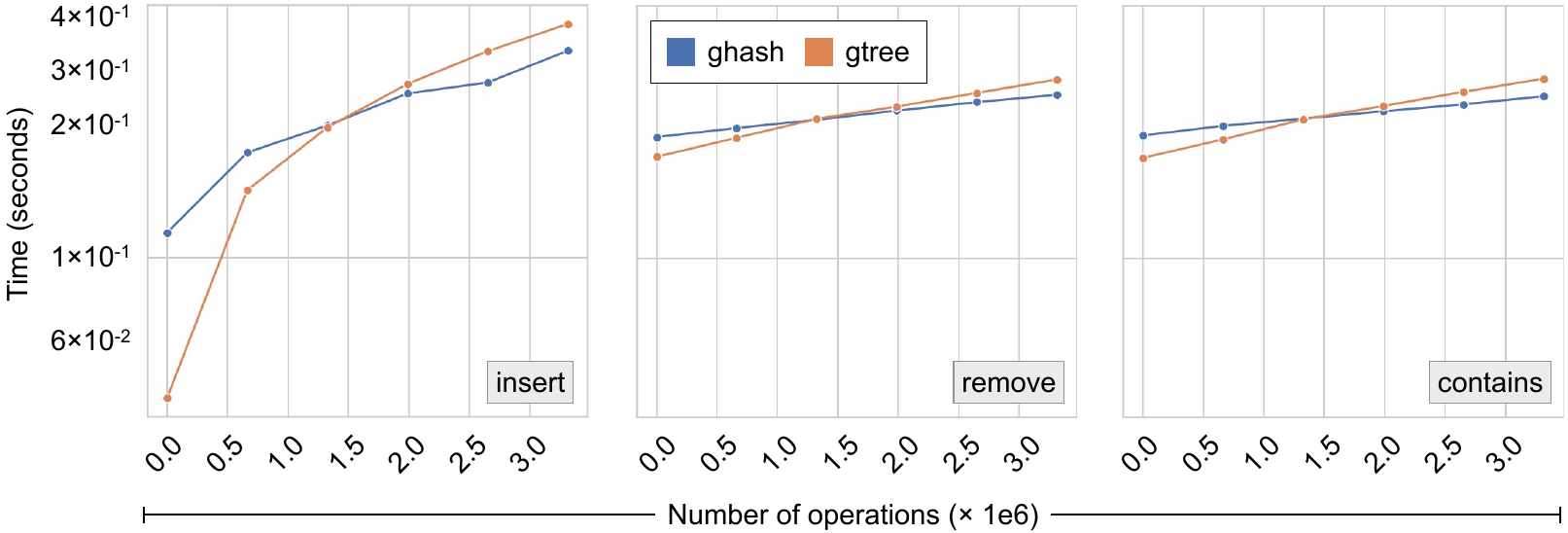}
\caption{Comparison of balanced trees (\textsc{GTree}) and hash tables (\textsc{GHashTable}) from \textsc{GLib}.}
\label{fig_treeVsHashDataStructures}
\end{figure}

At the operation scales shown in Figure~\ref{fig_allDataStructures}, \textsc{GTree} is consistently faster than \textsc{GHashTable}. However, when the workload increases further (we repeated the comparison for larger parameter values) \textsc{GHashTable} eventually overtakes \textsc{GTree} (see Figure~\ref{fig_treeVsHashDataStructures}). This crossover follows from the implementation trade-offs: balanced trees provide $O(\log n)$ guarantees, while hash tables typically offer amortized $O(1)$ behavior but pay higher constant costs for hashing, bucket management, and collision handling. For smaller workloads the hash-table constants dominate, favoring \textsc{GTree}; for larger workloads the different asymptotic growth rates drive the hash table to superior performance. Thus, the observed crossover reflects the interplay between constant-factor overheads and asymptotic scaling in these container implementations.

\subsection{CS7: Comparison with \textsc{CSmith}}
\label{sub_csmith}

Several programming language fuzzers enjoy widespread popularity.
The main goal of such tools, including \textsc{CSmith}~\cite{Yang11} and \textbf{YarpGen}~\cite{Livinskii20} (for C) or \textsc{ChiGen}~\cite{Vieira25} (for Verilog), is to uncover bugs in language processing systems such as compilers or interpreters.
The purpose of \benchgen{}, however, is different: it was designed to evaluate the performance of these systems.
Its ability to expose bugs is limited, since the programs it produces are less diverse than those generated by a fuzzer like \textsc{CSmith}.
On the other hand, \benchgen{} can stress-test compiler performance in ways that \textsc{CSmith} and related tools cannot.
This section illustrates this point through the following challenge: generate a collection of programs whose execution time is comparable to the average runtime of \textsc{Spec Cpu2017} programs, and whose behavior under optimization resembles that of real-world workloads.

\paragraph{Discussion. }
To compare \textsc{CSmith} and \benchgen{} under this challenge, we have used these two program generators to produce a collection of four benchmarks and compare them with the programs from \textsc{Spec Cpu2017}.
The adopted methodology is described below:
\begin{enumerate}
\item We ran the 28 \textsc{Spec Cpu2017} programs that \texttt{clang}/LLVM 20.1 can compile.
\item Using \textsc{CSmith}, we generated 10,000 programs and selected the four whose execution time was closest to the average runtime of the 28 \textsc{Spec Cpu2017} programs compiled without optimizations.
\item Using the four grammars described in Section~\ref{sub_gcc_vs_clang}, we generated four \benchgen{} programs whose runtime approximated the same average.
\item We then optimized each collection of benchmarks (28, 4, and 4 programs, respectively) under different optimization flags.
\end{enumerate}

\begin{table*}[t]
\centering
\begin{tabularx}{\textwidth}{lXXXX}
\toprule
 & \textbf{-O0} & \textbf{-O1} & \textbf{-O2} & \textbf{-O3} \\
\midrule
\textbf{BenchGen} & & & & \\
Cycles (Mean) & 1.44E+11 & 5.21E+10 & 5.20E+10 & 5.14E+10 \\
Ratio          & 1        & 2.76     & 2.77     & 2.80     \\
\midrule
\textbf{SPEC C2017} & & & & \\
Cycles (Mean) & 5.63E+11 & 2.43E+11 & 2.37E+11 & 2.30E+11 \\
Ratio          & 1        & 2.32     & 2.37     & 2.44     \\
\midrule
\textbf{CSmith} & & & & \\
Cycles (Mean) & 2.85E+06 & 1.57E+06 & 1.49E+06 & 1.44E+06 \\
Ratio          & 1        & 1.85     & 1.91     & 1.98     \\
\bottomrule
\end{tabularx}
\caption{Mean cycles and ratios for different benchmarks across optimization levels.}
\label{tab_vs_csmith}
\end{table*}

Table~\ref{tab_vs_csmith} presents the results of this experiment.
We stopped \benchgen{} as soon as we obtained programs whose running time was above the average \textsc{Spec} time, and chose the generation immediately before.
However, it is not possible to steer \textsc{CSmith} in such a way.
In other words, users cannot control \textsc{CSmith} to produce programs that execute within a given time frame.
Most \textsc{CSmith} programs tend to run for only a very short time.
Moreover, the impact of optimizations on these programs differs substantially from what we observe in real-world benchmarks such as those in the \textsc{Spec} collection, whereas the effects observed on \benchgen{} programs more closely resemble the behavior of genuine workloads.
This difference between the effects of compiler optimizations on \textsc{CSmith} and real-world programs had been noticed in previous work~\cite{Faustino21}.

\section{Related Work}
\label{sec_rw}

\benchgen{} is a tool designed to synthesize benchmarks.  
A number of works in the compiler literature have also proposed techniques for benchmark generation.  
In what follows, we review this body of work, emphasizing aspects that make \benchgen{} particularly well-suited for performance analysis.  

\paragraph{Compiler Fuzzers. }
The key distinction between compiler fuzzers and \benchgen{} lies in their primary objectives.  
Most existing tools are designed to uncover bugs in language processing systems, whereas \benchgen{} is specifically aimed at analyzing their performance.  
Consequently, \benchgen{} would be less effective than a fuzzer such as \textsc{CSmith} for identifying errors in C compilers.  
On the other hand, it enables tasks that are not easily supported by traditional fuzzers, such as studying the asymptotic behavior of compiler optimizations, comparing compilers in terms of compilation time, code size, and execution speed, or analyzing the evolution of a single compiler across multiple versions with respect to these same metrics.

Several tools are able to generate random strings from the production rules of a context-free grammar~\cite{Hodovan18,Bendrissou25,Vieira25,Zamudio24}. 
Indeed, the core ideas behind grammar-guided fuzzing date back to the 1970s~\cite{Purdom72}. 
A well-known limitation of this approach, however, is that for programming languages it is generally difficult to guarantee that the generated strings correspond to compilable programs. 
Although grammar-based fuzzers can enforce syntactic correctness, successful compilation typically depends on semantic constraints that exceed the expressive power of context-free grammars.

Even fuzzers explicitly designed to generate executable programs may fail to produce compilable code.
For example, Ricardo \textit{et al.}~\cite{Ricardo25} report a compilation success rate of 78.49\% when using \textsc{YarpGen}~\cite{Livinskii20} to generate C/C++ programs, while Vieira \textit{et al.} report a success rate of approximately 60\% when using the \textsc{ChiGen} fuzzer to produce Verilog designs.
This limitation is consistent with our experience in developing \benchgen, where a substantial portion of the engineering effort was devoted to ensuring that every generated program compiles without warnings and executes without runtime exceptions.

\paragraph{Benchmark Synthesizers. }
The development of compilers requires \textit{benchmarks}.
For this reason, some of the most celebrated papers in programming languages describe benchmark suites, such as \textsc{Spec CPU2006}~\cite{Henning06}, \textsc{MiBench}~\cite{Guthaus01}, \textsc{Rodinia}~\cite{Che09}, etc.
These benchmarks are manually curated and usually comprise a small number of programs.
A few years ago, Cummins et al.~\cite{Cummins17} demonstrated that this small size fails to cover the space of program features that a compiler is likely to explore during its lifetime.

The generation of benchmarks for tuning predictive compilers has been an active research field over the last ten years.
Early efforts aimed at the development of predictive optimizers used synthetic benchmarks designed to find compiler bugs.
Examples of such synthesizers include \textsc{CSmith}~\cite{Yang11}, \textsc{LDRGen}~\cite{Barany17}, and \textsc{Orange3}~\cite{Nagai14,Nakamura15}.
Although conceived as test case generators, these tools have also been used to improve the quality of optimized code emitted by mainstream C compilers~\cite{Hashimoto16,Barany18}.
Even \textsc{CompilerGym}~\cite{cummins22}, a tool for applying machine-learning-based code optimization techniques, provides randomly produced \textsc{CSmith} programs.
However, more recent developments indicate that synthetic codes tend to poorly reflect the behavior of programs written by humans; consequently, they produce deficient training sets~\cite{Faustino21,Goens19}.
Section~\ref{sub_csmith} discusses precisely these issues with \textsc{CSmith}: the difficulty of using it to generate programs that run for a reasonable amount of time and the ease with which compilers optimize such programs.

In order to produce more realistic benchmarks, a vast literature on extracting programs from repositories has recently emerged, seeking to build benchmark suites for training compilers.
Some of these works aim to generate benchmarks to feed machine learning models~\cite{Fowkes16,Gousios17,Palacio23}, for example.
This literature has some shortcomings when compared to \benchgen{}:
\begin{itemize}
\item Very large benchmark suites, such as \textsc{AnghaBench}~\cite{Faustino21} or \textsc{Poj}~\cite{Mou16}, compile but do not run.
\item Benchmark suites designed to run sometimes exhibit undefined behavior, such as \textsc{ExeBench}~\cite{Armengol22} and \textsc{ColaGen}~\cite{Berezov22}.
\item Techniques that generate benchmarks using LLMs, such as the system of Italiano and Cummins~\cite{Italiano25}, fail to produce large programs and take a long time to generate even a small number of programs, as recently shown by Guimar\~{a}es {\it et al}~\cite{Guimaraes25}.
\item The \textsc{Jotai} collection~\cite{Kind22}, which was built from programs that run without undefined behavior, contains only very small programs.
In our attempt to run the programs available in \textsc{CompilerGym}~\cite{cummins22}, the longest-running program took only 0.017 seconds when compiled with \texttt{clang -O0}.
\end{itemize}

\section{Conclusion}
\label{sec_conc}

This paper has presented a methodology for generating benchmarks based on the observation that programs often exhibit self-similar structures, and thus can be described as derivations of L-Systems.
We validated this idea through the design and implementation of \benchgen, a tool capable of producing benchmarks in multiple programming languages.
In contrast to typical fuzzers, \benchgen{} is multilingual and allows users to generate programs of arbitrary size whose execution exercises complex control-flow patterns.

\paragraph{Future Work. }
Beyond the concrete implementation of \benchgen, the key contribution of this paper is the proposal that programs can be systematically generated as instances of L-Systems.
The particular L-System flavor explored here employs a combination of seven constructs (three control-flow structures and four data-structure behaviors).
Nevertheless, the approach can be extended to a much broader set of syntactic features.
For example, the current version of \benchgen{} generates programs restricted to a single data structure at a time, whereas nothing in the formalism prevents richer combinations of data structures.
Likewise, additional control-flow constructs such as \texttt{switch}, \texttt{do-while}, or even \texttt{goto} could be naturally incorporated.
We leave the investigation of such extensions as promising directions for future work.

\section*{Data Availability Statement}

\benchgen{} is available \url{https://github.com/lac-dcc/BenchGen}, under the Apache-2.0 license.

\section*{Acknowledgment}
This project was supported by Google and by FAPEMIG (Grant APQ-00440-23).  
We are grateful to Xinliang (David) Li and Victor Lee for their efforts in making the Google sponsorship possible.  
We also thank Lucas Victor Silva for producing the data in Table~\ref{tab_vs_csmith}, and Bruno Pena Ba\^{e}ta and Matheus Alc\^{a}ntara for their work on an earlier version of \benchgen.

\bibliographystyle{plain}
\bibliography{ref}

@inproceedings{Hodovan18,
author = {Hodov\'{a}n, Ren\'{a}ta and Kiss, \'{A}kos and Gyim\'{o}thy, Tibor},
title = {Grammarinator: a grammar-based open source fuzzer},
year = {2018},
isbn = {9781450360531},
publisher = {Association for Computing Machinery},
address = {New York, NY, USA},
url = {https://doi.org/10.1145/3278186.3278193},
doi = {10.1145/3278186.3278193},
booktitle = {Proceedings of the 9th ACM SIGSOFT International Workshop on Automating TEST Case Design, Selection, and Evaluation},
pages = {45–48},
numpages = {4},
location = {Lake Buena Vista, FL, USA},
series = {A-TEST 2018}
}

@article{Bendrissou25,
author = {Bendrissou, Bachir and Cadar, Cristian and Donaldson, Alastair F.},
title = {Grammar Mutation for Testing Input Parsers},
year = {2025},
issue_date = {May 2025},
publisher = {Association for Computing Machinery},
address = {New York, NY, USA},
volume = {34},
number = {4},
issn = {1049-331X},
url = {https://doi.org/10.1145/3708517},
doi = {10.1145/3708517},
journal = {ACM Trans. Softw. Eng. Methodol.},
month = apr,
articleno = {116},
numpages = {21},
}

@inproceedings{Zamudio24,
author = {Zamudio Amaya, Jos\'{e} Antonio},
title = {Shaping Test Inputs in Grammar-Based Fuzzing},
year = {2024},
isbn = {9798400706127},
publisher = {Association for Computing Machinery},
address = {New York, NY, USA},
url = {https://doi.org/10.1145/3650212.3685553},
doi = {10.1145/3650212.3685553},
booktitle = {ISSTA},
pages = {1901–1905},
numpages = {5},
location = {Vienna, Austria},
}

@article{Purdom72,
author = {Purdom, Paul},
title = {A sentence generator for testing parsers},
year = {1972},
issue_date = {Sep 1972},
publisher = {BIT Computer Science and Numerical Mathematics},
address = {USA},
volume = {12},
number = {3},
issn = {0006-3835},
url = {https://doi.org/10.1007/BF01932308},
doi = {10.1007/BF01932308},
journal = {BIT},
month = sep,
pages = {366–375},
numpages = {10},
}

@inproceedings{Guimaraes25,
 author = {Gabriel Ricardo and Natanael Santos Junior and Flavio Figueiredo and Fernando Pereira},
 title = { On the Practicality of LLM-Based Compiler Fuzzing},
 booktitle = {SBLP},
 location = {Recife/PE},
 year = {2025},
 pages = {59--66},
 publisher = {SBC},
 address = {Porto Alegre, RS, Brasil},
 doi = {10.5753/sblp.2025.12264},
 url = {https://sol.sbc.org.br/index.php/sblp/article/view/36949}
}

@article{Ellison12,
author = {Ellison, Chucky and Rosu, Grigore},
title = {An executable formal semantics of C with applications},
year = {2012},
issue_date = {January 2012},
publisher = {Association for Computing Machinery},
address = {New York, NY, USA},
volume = {47},
number = {1},
issn = {0362-1340},
url = {https://doi.org/10.1145/2103621.2103719},
doi = {10.1145/2103621.2103719},
journal = {SIGPLAN Not.},
month = jan,
pages = {533–544},
numpages = {12}
}

@article{Baudin21,
author = {Baudin, Patrick and Bobot, Fran\c{c}ois and B\"{u}hler, David and Correnson, Lo\"{\i}c and Kirchner, Florent and Kosmatov, Nikolai and Maroneze, Andr\'{e} and Perrelle, Valentin and Prevosto, Virgile and Signoles, Julien and Williams, Nicky},
title = {The dogged pursuit of bug-free C programs: the Frama-C software analysis platform},
year = {2021},
issue_date = {August 2021},
publisher = {Association for Computing Machinery},
address = {New York, NY, USA},
volume = {64},
number = {8},
issn = {0001-0782},
url = {https://doi.org/10.1145/3470569},
doi = {10.1145/3470569},
journal = {Commun. ACM},
month = jul,
pages = {56–68},
numpages = {13}
}

@inproceedings{Nethercote07,
author = {Nethercote, Nicholas and Seward, Julian},
title = {Valgrind: a framework for heavyweight dynamic binary instrumentation},
year = {2007},
isbn = {9781595936332},
publisher = {Association for Computing Machinery},
address = {New York, NY, USA},
url = {https://doi.org/10.1145/1250734.1250746},
doi = {10.1145/1250734.1250746},
booktitle = {PLDI},
pages = {89–100},
numpages = {12},
location = {San Diego, California, USA},
}

@inproceedings{Serebryany12,
author = {Serebryany, Konstantin and Bruening, Derek and Potapenko, Alexander and Vyukov, Dmitry},
title = {AddressSanitizer: a fast address sanity checker},
year = {2012},
publisher = {USENIX Association},
address = {USA},
booktitle = {USENIX ATC},
pages = {28},
numpages = {1},
location = {Boston, MA},
}

@inproceedings{Dehao16,
author = {Chen, Dehao and Li, David Xinliang and Moseley, Tipp},
title = {AutoFDO: automatic feedback-directed optimization for warehouse-scale applications},
year = {2016},
isbn = {9781450337786},
publisher = {Association for Computing Machinery},
address = {New York, NY, USA},
url = {https://doi.org/10.1145/2854038.2854044},
doi = {10.1145/2854038.2854044},
booktitle = {CGO},
pages = {12–23},
numpages = {12},
location = {Barcelona, Spain},
}

@article{Moreira21,
author = {Moreira, Ang\'{e}lica Aparecida and Ottoni, Guilherme and Quint\~{a}o Pereira, Fernando Magno},
title = {VESPA: static profiling for binary optimization},
year = {2021},
issue_date = {October 2021},
publisher = {Association for Computing Machinery},
address = {New York, NY, USA},
volume = {5},
number = {OOPSLA},
url = {https://doi.org/10.1145/3485521},
doi = {10.1145/3485521},
journal = {Proc. ACM Program. Lang.},
month = oct,
articleno = {144},
numpages = {28}
}

@inproceedings{Panchenko19,
author = {Panchenko, Maksim and Auler, Rafael and Nell, Bill and Ottoni, Guilherme},
title = {BOLT: a practical binary optimizer for data centers and beyond},
year = {2019},
isbn = {9781728114361},
publisher = {IEEE Press},
booktitle = {CGO},
pages = {2–14},
numpages = {13},
location = {Washington, DC, USA},
}

@book{Strecansky20,
  title={Hands-On High Performance with Go: Boost and optimize the performance of your Golang applications at scale with resilience},
  author={Strecansky, Bob},
  year={2020},
  publisher={Packt Publishing Ltd}
}

@article{Bezanson18,
author = {Bezanson, Jeff and Chen, Jiahao and Chung, Benjamin and Karpinski, Stefan and Shah, Viral B. and Vitek, Jan and Zoubritzky, Lionel},
title = {Julia: dynamism and performance reconciled by design},
year = {2018},
issue_date = {November 2018},
publisher = {Association for Computing Machinery},
address = {New York, NY, USA},
volume = {2},
number = {OOPSLA},
url = {https://doi.org/10.1145/3276490},
doi = {10.1145/3276490},
journal = {Proc. ACM Program. Lang.},
month = oct,
articleno = {120},
numpages = {23}
}

@inproceedings{Olivo15,
author = {Olivo, Oswaldo and Dillig, Isil and Lin, Calvin},
title = {Static detection of asymptotic performance bugs in collection traversals},
year = {2015},
isbn = {9781450334686},
publisher = {Association for Computing Machinery},
address = {New York, NY, USA},
url = {https://doi.org/10.1145/2737924.2737966},
doi = {10.1145/2737924.2737966},
booktitle = {PLDI},
pages = {369–378},
numpages = {10},
location = {Portland, OR, USA},
}

@inproceedings{Han16,
author = {Han, Xue and Yu, Tingting},
title = {An Empirical Study on Performance Bugs for Highly Configurable Software Systems},
year = {2016},
isbn = {9781450344272},
publisher = {Association for Computing Machinery},
address = {New York, NY, USA},
url = {https://doi.org/10.1145/2961111.2962602},
doi = {10.1145/2961111.2962602},
booktitle = {ESEM},
articleno = {23},
numpages = {10},
location = {Ciudad Real, Spain},
}

@inproceedings{Curtsinger15,
author = {Curtsinger, Charlie and Berger, Emery D.},
title = {Coz: finding code that counts with causal profiling},
year = {2015},
isbn = {9781450338349},
publisher = {Association for Computing Machinery},
address = {New York, NY, USA},
url = {https://doi.org/10.1145/2815400.2815409},
doi = {10.1145/2815400.2815409},
booktitle = {SOSP},
pages = {184–197},
numpages = {14},
location = {Monterey, California},
}

@misc{Vieira25,
      title={Bottom-Up Generation of Verilog Designs for Testing EDA Tools}, 
      author={João Victor Amorim Vieira and Luiza de Melo Gomes and Rafael Sumitani and Raissa Maciel and Augusto Mafra and Mirlaine Crepalde and Fernando Magno Quintão Pereira},
      year={2025},
      eprint={2504.06295},
      archivePrefix={arXiv},
      primaryClass={cs.AR},
      url={https://arxiv.org/abs/2504.06295}, 
}

@techreport{Kind22,
  title={Jotai: a methodology for the generation of executable C benchmarks},
  author={Kind, Cecilia Conde and Canesche, Michael and Pereira, Fernando Magno Quintao},
  year={2022},
  institution={Technical Report 02-2022. Universidade Federal de Minas Gerais}
}

@inproceedings{Italiano25,
author = {Italiano, Davide and Cummins, Chris},
title = {Finding Missed Code Size Optimizations in Compilers using Large Language Models},
year = {2025},
isbn = {9798400714078},
publisher = {Association for Computing Machinery},
address = {New York, NY, USA},
url = {https://doi.org/10.1145/3708493.3712686},
doi = {10.1145/3708493.3712686},
booktitle = {International Conference on Compiler Construction},
pages = {81–91},
numpages = {11},
location = {Las Vegas, NV, USA},
}

@inproceedings{Armengol22,
author = {Armengol-Estap\'{e}, Jordi and Woodruff, Jackson and Brauckmann, Alexander and Magalh\~{a}es, Jos\'{e} Wesley de Souza and O'Boyle, Michael F. P.},
title = {ExeBench: an ML-scale dataset of executable C functions},
year = {2022},
isbn = {9781450392730},
publisher = {Association for Computing Machinery},
address = {New York, NY, USA},
url = {https://doi.org/10.1145/3520312.3534867},
doi = {10.1145/3520312.3534867},
booktitle = {MAPS},
pages = {50–59},
numpages = {10},
location = {San Diego, CA, USA},
}

@inproceedings{Mou16,
author = {Mou, Lili and Li, Ge and Zhang, Lu and Wang, Tao and Jin, Zhi},
title = {Convolutional Neural Networks over Tree Structures for Programming Language Processing},
year = {2016},
publisher = {AAAI Press},
address = {Palo Alto, CA, US},
booktitle = {AAAI},
pages = {1287–1293},
}

@inproceedings{Kitaura18,
author = {Kitaura, Kota and Ishiura, Nagisa},
title = {Random testing of compilers’ performance based on mixed static and dynamic code comparison},
year = {2018},
isbn = {9781450360531},
publisher = {Association for Computing Machinery},
address = {New York, NY, USA},
url = {https://doi.org/10.1145/3278186.3278192},
doi = {10.1145/3278186.3278192},
booktitle = {Proceedings of the 9th ACM SIGSOFT International Workshop on Automating TEST Case Design, Selection, and Evaluation},
pages = {38–44},
numpages = {7},
location = {Lake Buena Vista, FL, USA},
series = {A-TEST 2018}
}

@ARTICLE{Manes21,
  author={Manes, Valentin J.M. and Han, HyungSeok and Han, Choongwoo and Cha, Sang Kil and Egele, Manuel and Schwartz, Edward J. and Woo, Maverick},
  journal={IEEE Transactions on Software Engineering}, 
  title={The Art, Science, and Engineering of Fuzzing: A Survey}, 
  year={2021},
  volume={47},
  number={11},
  pages={2312-2331},
  doi={10.1109/TSE.2019.2946563}
}

@article{Wang18,
  author    = {Zheng Wang and
    Michael F. P. O'Boyle},
  title     = {Machine Learning in Compiler Optimization},
  journal   = {Proceedings of the {IEEE}},
  volume    = {106},
  number    = {11},
  pages     = {1879--1901},
  year      = {2018},
  doi       = {10.1109/JPROC.2018.2817118}
}

@inproceedings{Yang11,
author = {Yang, Xuejun and Chen, Yang and Eide, Eric and Regehr, John},
title = {Finding and understanding bugs in C compilers},
year = {2011},
isbn = {9781450306638},
publisher = {Association for Computing Machinery},
address = {New York, NY, USA},
url = {https://doi.org/10.1145/1993498.1993532},
doi = {10.1145/1993498.1993532},
booktitle = {Proceedings of the 32nd ACM SIGPLAN Conference on Programming Language Design and Implementation},
pages = {283–294},
numpages = {12},
location = {San Jose, California, USA},
series = {PLDI '11}
}

@article{Lindenmayer68,
  title={Mathematical models for cellular interactions in development I. Filaments with one-sided inputs},
  author={Lindenmayer, Aristid},
  journal={Journal of theoretical biology},
  volume={18},
  number={3},
  pages={280--299},
  year={1968},
  publisher={Elsevier}
}

@article{Livinskii20,
author = {Livinskii, Vsevolod and Babokin, Dmitry and Regehr, John},
title = {Random testing for C and C++ compilers with YARPGen},
year = {2020},
issue_date = {November 2020},
publisher = {Association for Computing Machinery},
address = {New York, NY, USA},
volume = {4},
number = {OOPSLA},
url = {https://doi.org/10.1145/3428264},
doi = {10.1145/3428264},
journal = {Proc. ACM Program. Lang.},
month = nov,
articleno = {196},
numpages = {25}
}

@inproceedings{Xia24,
author = {Xia, Chunqiu Steven and Paltenghi, Matteo and Le Tian, Jia and Pradel, Michael and Zhang, Lingming},
title = {Fuzz4All: Universal Fuzzing with Large Language Models},
year = {2024},
isbn = {9798400702174},
publisher = {Association for Computing Machinery},
address = {New York, NY, USA},
url = {https://doi.org/10.1145/3597503.3639121},
doi = {10.1145/3597503.3639121},
booktitle = {Proceedings of the IEEE/ACM 46th International Conference on Software Engineering},
articleno = {126},
numpages = {13},
location = {Lisbon, Portugal},
series = {ICSE '24}
}

@inproceedings{Barany17,
  author    = {Gerg{\"{o}} Barany},
  title     = {Liveness-Driven Random Program Generation},
  booktitle = {LOPSTR},
  pages     = {112--127},
  year      = {2017},
  doi       = {10.1007/978-3-319-94460-9\_7},
  publisher = {Springer},
  address   = {Heidelberg, Germany},
}

@inproceedings{Barany18,
 author = {Barany, Gerg\"{o}},
 title = {Finding Missed Compiler Optimizations by Differential Testing},
 booktitle = {Proceedings of the 27th International Conference on Compiler Construction},
 series = {CC 2018},
 year = {2018},
 isbn = {978-1-4503-5644-2},
 location = {Vienna, Austria},
 pages = {82--92},
 numpages = {11},
 doi = {10.1145/3178372.3179521},
 acmid = {3179521},
 publisher = {ACM},
 address = {New York, NY, USA},
}

@InProceedings{Berezov22,
  author =	{Berezov, Maksim and Ancourt, Corinne and Zawalska, Justyna and Savchenko, Maryna},
  title =	{{COLA-Gen: Active Learning Techniques for Automatic Code Generation of Benchmarks}},
  booktitle =	{PARMA-DITAM},
  pages =	{3:1--3:14},
  series =	{Open Access Series in Informatics (OASIcs)},
  ISBN =	{978-3-95977-231-0},
  ISSN =	{2190-6807},
  year =	{2022},
  volume =	{100},
  editor =	{Palumbo, Francesca and Bispo, Jo\~{a}o and Cherubin, Stefano},
  publisher =	{Schloss Dagstuhl -- Leibniz-Zentrum f{\"u}r Informatik},
  address =	{Dagstuhl, Germany},
  URL =		{https://drops.dagstuhl.de/opus/volltexte/2022/16119},
  doi =		{10.4230/OASIcs.PARMA-DITAM.2022.3}
}

@inproceedings{Che09,
 author = {Che, Shuai and Boyer, Michael and Meng, Jiayuan and Tarjan, David and Sheaffer, Jeremy W. and Lee, Sang-Ha and Skadron, Kevin},
 title = {Rodinia: A Benchmark Suite for Heterogeneous Computing},
 booktitle = {IISWC},
 year = {2009},
 isbn = {978-1-4244-5156-2},
 pages = {44--54},
 doi = {10.1109/IISWC.2009.5306797},
 publisher = {IEEE},
 address = {Washington, DC, USA},
}

@inproceedings{Cummins17,
  author = {Cummins, Chris and Petoumenos, Pavlos and Wang, Zheng and Leather, Hugh},
  title = {Synthesizing Benchmarks for Predictive Modeling},
  booktitle = {CGO},
  year = {2017},
  pages = {86--99},
  publisher = {IEEE},
  doi = {10.1109/CGO.2017.7863731},
  address = {Piscataway, NJ, USA},
}

@inproceedings{Faustino21,
  author    = {Anderson Faustino da Silva and
               Bruno Conde Kind and
               Jos{\'{e}} Wesley de Souza Magalh{\~{a}}es and
               Jer{\^{o}}nimo Nunes Rocha and
               Breno Campos Ferreira Guimar{\~{a}}es and
               Fernando Magno Quint{\~{a}}o Pereira},
  title     = {{AnghaBench:} {A} Suite with One Million Compilable {C} Benchmarks
               for Code-Size Reduction},
  booktitle = {CGO},
  pages     = {378--390},
  year      = {2021},
  publisher = {{IEEE}},
  address = {Los Alamitos, CA, USA},
  doi       = {10.1109/CGO51591.2021.9370322},
}

@inproceedings{Fowkes16,
author = {Fowkes, Jaroslav and Sutton, Charles},
title = {Parameter-Free Probabilistic API Mining across GitHub},
year = {2016},
isbn = {9781450342186},
publisher = {Association for Computing Machinery},
address = {New York, NY, USA},
url = {https://doi.org/10.1145/2950290.2950319},
doi = {10.1145/2950290.2950319},
booktitle = {FSE},
pages = {254–265},
numpages = {12},
location = {Seattle, WA, USA},
}

@inproceedings{Gousios17,
author = {Gousios, Georgios and Spinellis, Diomidis},
title = {Mining Software Engineering Data from GitHub},
year = {2017},
isbn = {9781538615898},
publisher = {IEEE Press},
address = {Washington, DC, US},
url = {https://doi.org/10.1109/ICSE-C.2017.164},
doi = {10.1109/ICSE-C.2017.164},
booktitle = {ICSE-C},
pages = {501–502},
numpages = {2},
location = {Buenos Aires, Argentina},
}

@inproceedings{Guthaus01,
 author = {Guthaus, M. R. and Ringenberg, J. S. and Ernst, D. and Austin, T. M. and Mudge, T. and Brown, R. B.},
 title = {{MiBench}: A Free, Commercially Representative Embedded Benchmark Suite},
 booktitle = {WWC},
 year = {2001},
 isbn = {0-7803-7315-4},
 pages = {3--14},
 doi = {10.1109/WWC.2001.15},
 publisher = {IEEE},
 address = {Washington, DC, USA},
}

@article{Hashimoto16,
  title={Detecting arithmetic optimization opportunities for C compilers by randomly generated equivalent programs},
  author={Hashimoto, Atsushi and Ishiura, Nagisa},
  journal={IPSJ Transactions on System LSI Design Methodology},
  volume={9},
  pages={21--29},
  year={2016},
  publisher={Information Processing Society of Japan}
}

@article{Henning06,
 author = {Henning, John L.},
 title = {{SPEC} {CPU2006} Benchmark Descriptions},
 journal = {SIGARCH Comput. Archit. News},
 volume = {34},
 number = {4},
 month = sep,
 year = {2006},
 issn = {0163-5964},
 pages = {1--17},
 doi = {10.1145/1186736.1186737},
 publisher = {ACM},
 address = {New York, NY, USA},
}

@article{Nagai14,
  title={Reinforcing Random Testing of Arithmetic Optimization of C Compilers by Scaling up Size and Number of Expressions},
  author={Eriko Nagai and Atsushi Hashimoto and Nagisa Ishiura},
  journal={IPSJ Trans. System LSI Design Methodology},
  year={2014},
  volume={7},
  pages={91-100}
}

@misc{Nakamura15,
  title={Introducing Loop Statements in Random Testing of C compilers Based on Expected Value Calculation},
  author={Nakamura, Kazuhiro and Ishiura, Nagisa},
  booktitle={SASIMI},
  pages={226--227},
  year={2015}
}

@misc{Palacio23,
      title={Evaluating and Explaining Large Language Models for Code Using Syntactic Structures},
      author={David N Palacio and Alejandro Velasco and Daniel Rodriguez-Cardenas and Kevin Moran and Denys Poshyvanyk},
      year={2023},
      eprint={2308.03873},
      archivePrefix={arXiv},
      primaryClass={cs.SE}
}

@inproceedings{Goens19,
 author = {Goens, Andr{\'e}s and Brauckmann, Alexander and Ertel, Sebastian and Cummins, Chris and Leather, Hugh and Castrillon, Jeronimo},
 title = {A Case Study on Machine Learning for Synthesizing Benchmarks},
 booktitle = {Proceedings of the 3rd ACM SIGPLAN International Workshop on Machine Learning and Programming Languages},
 series = {MAPL 2019},
 year = {2019},
 isbn = {978-1-4503-6719-6},
 location = {Phoenix, AZ, USA},
 pages = {38--46},
 numpages = {9},
 doi = {10.1145/3315508.3329976},
 acmid = {3329976},
 publisher = {ACM},
 address = {New York, NY, USA},
}

@inproceedings{cummins22,
  title={Compilergym: Robust, performant compiler optimization environments for ai research},
  author={Cummins, Chris and Wasti, Bram and Guo, Jiadong and Cui, Brandon and Ansel, Jason and Gomez, Sahir and Jain, Somya and Liu, Jia and Teytaud, Olivier and Steiner, Benoit and others},
  booktitle={CGO},
  pages={92--105},
  year={2022},
  publisher={IEEE},
  address = {New York, USA}
}

@inproceedings{Ricardo25,
 author = {Gabriel Ricardo and Natanael Santos Junior and Flavio Figueiredo and Fernando Pereira},
 title = { On the Practicality of LLM-Based Compiler Fuzzing},
 booktitle = {SBLP},
 location = {Recife/PE},
 year = {2025},
 issn = {0000-0000},
 pages = {59--66},
 publisher = {SBC},
 address = {Porto Alegre, RS, Brasil},
 doi = {10.5753/sblp.2025.12264},
 url = {https://sol.sbc.org.br/index.php/sblp/article/view/36949}
}

\end{document}